\documentclass[lettersize,journal]{IEEEtran}
\usepackage{amsmath,amsfonts}
\usepackage{algorithmic}
\usepackage{algorithm}
\usepackage{array}
\usepackage[caption=false,font=normalsize,labelfont=sf,textfont=sf]{subfig}
\usepackage{textcomp}
\usepackage{makecell}
\usepackage{stfloats}
\usepackage{booktabs} 
\usepackage{multirow}
\usepackage{url}
\usepackage{verbatim}
\usepackage{graphicx}
\usepackage{ulem}
\usepackage{multirow}
\usepackage{booktabs}
\usepackage{hyperref}
\hypersetup{
	colorlinks=true,
	allcolors=blue,
	pdfstartview=Fit,
	breaklinks=true}
\usepackage{float}
\newcommand{\rev}[1]{\textcolor{black}{#1}}
\begin{document}
\title{DBHN-Net: Dual-Branch Hybrid Neural Network For Low-Complexity Monaural Speech Enhancement}
\author{Cunhang Fan,~\IEEEmembership{Member,~IEEE,} Enrui Liu,  Jing Zhou, Jian Kang, Jie Li, Andong Li,~\IEEEmembership{Member,~IEEE,} \\Jian Zhou, Zhao Lv,~\IEEEmembership{Member,~IEEE,} Xuelong Li,~\IEEEmembership{Fellow,~IEEE}

\thanks{This work is supported by the {Brain Science and Brain-like Intelligence Technology-National Science and Technology Major Project (No.2021ZD0201500)}, the National Natural Science Foundation of China (NSFC)(No.62571002, 62476004), Excellent Youth Foundation of Anhui Scientific Committee (No.2408085Y034).(Corresponding author: Enrui Liu, Xuelong Li.)

Cunhang Fan,
Enrui Liu,
Jian Zhou and Zhao Lv are with State Key Laboratory of Opto-Electronic Information Acquisition and Protection Technology, (School of Computer Science and Technology), Anhui University, Hefei, 230601, Anhui, P. R. China (e-mail: e23201090@stu.ahu.edu.cn; cunhang.fan@ahu.edu.cn; jzhou@ahu.edu.cn; kjlz@ahu.edu.cn).

Jing Zhou,
Jian Kang and Jie Li are with China Telecom Artificial Intelligence Technology (Beijing) Co., Ltd (e-mail: zhouj100@chinatelecom.cn; kangj30@chinatelecom.cn; lij86@chinatelecom.cn;).

Andong Li is with Institute of Acoustics, University of Chinese Academy of Sciences, Beijing 100190, China (e-mail:liandong@mail.ioa.ac.cn).

Xuelong Li is with Institute of Artificial Intelligence (TeleAI), China Telecom, China  (xuelong\_li@ieee.org).

}}

\markboth{Journal of \LaTeX\ Class Files,~Vol.~14, No.~8, August~2021}%
{Shell \MakeLowercase{\textit{et al.}}: A Sample Article Using IEEEtran.cls for IEEE Journals}


\maketitle

\begin{abstract}
\rev{Although artificial neural network (ANN) based speech enhancement (SE) methods demonstrate excellent performance, the high computational complexity and high energy consumption hinder their deployment in practical front-end processing tasks.} Currently, the spiking neural networks (SNNs) have shown potential in reducing power consumption. However, the discrete binary activation and complex spatio-temporal dynamics of SNNs often result in information loss. The current challenge therefore focuses on how to maintain performance and reduce computational complexity. To address this issue, this work propose a Dual-Branch Hybrid Neural (DBHN) Network. 1) In terms of network architecture: A dual-branch network integrating ANN and SNN was designed, where the SNN branch reduces power consumption while the ANN branch addresses information loss; The BandSplit and Time-Frequency (TF) -Mamba modules were developed to simultaneously compress energy consumption and enhance model performance; Spiking Feature Extraction Group (SFEG) and  Information Transformation Block (ITB) components were implemented with residual connections to mitigate information loss while further refining feature representations. 2) To facilitate inter-branch information fusion: An Interaction module was designed to promote information exchange at various stages of the dual-branch network; A TF-Cross Attention-Fusion module was designed to perform time-frequency domain fusion of dual-branch information while data-adaptively guiding the SNN branch to retain more critical information. Results show that the proposed model maintains superior performance across three public datasets while achieving an average 7.5 fold reduction in computational complexity compared to baseline models.
\end{abstract}


\section{Introduction}
\IEEEPARstart{S}{p}eech enhancement (SE) is a critical task in audio signal processing, aiming to recover clean speech from noisy recordings \cite{Denoising,tpami1,bsdb}. This process is essential because background noise not only reduces listening clarity but also adversely affects downstream systems, including automatic speech recognition (ASR) \cite{xiangwang,WaveSpecEnc,c:1}. Furthermore, SE applications are becoming increasingly prevalent across real-world scenarios including smart devices, vehicular systems, and home automation. This widespread deployment highlights the urgen need for SE models that simultaneously maintain effectiveness while achieving low power consumption \cite{fan2025seeing,tpami2,liuenrui}.

\begin{table}[t]
\caption{\rev{Summary of the Characteristics of Different Networks. Here, "Arch." refers to Architecture, "Energy" refers to computational complexity and energy consumption, and "Per." refers to performance. A greater number of "+" indicates higher computational complexity and energy consumption, while a greater number of "*" indicates better performance.}}
\begin{center}
\setlength{\tabcolsep}{1pt} 
\begin{tabular}{c | c c c}
\toprule
\textbf{Arch.} & \textbf{\makecell{Energy}} & \textbf{Per.} & \textbf{Key Structural Characteristics}                           \\ \midrule
ANN                   & \makecell{+\\+\\+\\+\\+}                & \makecell{$*$\\$*$\\$*$\\$*$\\$*$}                & \makecell{Diverse approaches with specific focuses:\\• Dual-branch → Magnitude+Phase.\\• Fullband-subband → Wideband+Narrowband.\\• Intra-/Inter-block → Local+Global.\\• Pure ANN →  For various SE objectives.} \\ \midrule
ANN→SNN              & \makecell{+\\+}                   & \makecell{$*$\\$*$}                     & \makecell{•Maintains the reference: \\ANN's core architecture and design.\\ •Limited by information loss.}                                                                                                                                                  \\ \midrule
SNN    & +                    & \makecell{$*$}                   & \makecell{•Focus on SNN's information loss. \\• Balancing performance and energy.}                                                                                                                           \\ \midrule

\makecell{\textbf{Hybrid Networks}\\ \textbf{(OURS)}}              &\makecell{+\\+\\+}                 & \makecell{$*$\\$*$\\$*$\\$*$\\$*$}                 & \makecell{•Hybrid Network:\\ANN-SNN parallel architecture. \\ •Information fusion and interaction.\\•To reduce computational complexity.\\•To guarantee performance.}                                                                                                                                                                   \\ \bottomrule
\end{tabular}
\label{table1}
\end{center}
\end{table}

In the past years, Artificial Neural Network (ANN)-based SE methods have undergone rapid development \cite{ai1,wahab2024compact}. In terms of sequence modeling architectures, the field has evolved from initial Convolutional Neural Network (CNN) -based \cite{dccrn} approaches to Recurrent Neural Networks (RNNs) \cite{DPRNN} for capturing historical information, followed by Long Short-Term Memory (LSTM) \cite{LSTM} to address RNNs' historical information loss limitations. Subsequently, Temporal Convolutional Module (TCM) and Squeezed Temporal Convolutional Module (S-TCM) \cite{ctsnet} structures were incorporated into SE networks. To better handle long-term dependencies, transformer was introduced and demonstrated superior performance. More recently, Mamba \cite{bsdb} has been applied to SE systems to mitigate the quadratic complexity issues of the transformer \cite{DBTNET}. Beyond sequence modeling components, SE network architectures have also evolved significantly - progressing from single-stage models that only enhanced magnitude spectra while neglecting phase information \cite{CRN}, to those employing complex spectral processing upon recognizing the importance of phase information \cite{GCRN}. Subsequent research revealed that directly processing the complex spectrum could lead to ``compensation effects'' between phase and magnitude information \cite{compensation,zhanghongyu}. To mitigate this issue, decomposed multistage models employing parallel dual-branch \cite{gagnet} architectures have been proposed to extract phase and magnitude information separately. In addition to these structural innovations, both frequency-band segmentation strategies and full/sub-band \cite{fullsubnet} approaches have been introduced to focus on more critical frequency components. While these improvements have boosted the performance of SE networks, this achievement is at the expense of a substantial increase in model computational complexity and power consumption. As a front-end processing task, SE system must achieve low power consumption to enable practical deployment \cite{lowpower,wenti}.

Recently, there has been growing research interest in Spiking Neural Networks (SNNs), which are recognized for their unique brain-inspired computing mechanisms that promise ultra-low power consumption \cite{Spiking-FullSubNet,snnlow}. SNNs mimic the brain’s electrical signaling by transmitting information through activated neurons via spike streams \cite{aydin2024hsnnannlow}. This communication paradigm exhibits asynchronous and event-driven characteristics \cite{karamimanesh2025spiking}. Moreover, merely a minority of neurons within SNNs fire to produce spikes at any given time step, resulting in an uneven distribution of neuronal activity that inherently induces network sparsity. Recently, the Intel Neuromorphic Deep Noise Suppression Challenge solicited high-performance SNN models for SE tasks \cite{inteldnschallenge}. However, the SNN-based SE models still face several critical challenges. First, current SNN-based SE models primarily convert existing ANN architectures into spiking networks through activation function modifications \cite{snnannlow}, this often leads to structural redundancy and training difficulties due to inherent network incompatibilities with spike-based information processing \cite{zhong2025hynita}. Second, speech signals typically contain complex information exhibiting multi-scale temporal variations. The binary (0/1) nature of spike-based propagation in SNNs inevitably leads to information loss, which significantly degrades model performance \cite{snnlossin}.

Effectively integrating the distinct characteristics of both neural network types to achieve complementary advantages through architectural and informational fusion presents a significant challenge. To address this, a Dual-Branch Hybrid Neural Network (DBHN) comprising an ANN branch and an SNN branch is proposed, which is designed to maintain performance while minimizing computational complexity. The ANN branch incorporates a BandSplit module to focus on spectrally significant information while reducing computational overhead, along with a Mamba-based TF-Mamba block for efficient sequential modeling that further reduces complexity without performance degradation. The SNN branch utilizes Leaky Integrate-and-Fire (LIF) units to convert information into spike signals, featuring a novel LIF-based Spiking Feature Extraction Group (SFEG) that preserves critical information within limited representations, and an Information Transformation Block (ITB) that converts discrete spikes back to continuous signals at the final stage, refining information while mitigating binary discretization losses. Furthemore, to enable effective cross-branch fusion, we develop an Interaction module that operates throughout all processing stages to progressively integrate discrete and continuous representations, and a TF-Cross Attention-Fusion (TF-CAF) module that performs final integration through time-frequency dual-domain cross-attention, specifically optimized for audio signal characteristics.

The main contributions of this paper are as follows:
\begin{list}{}{}
\item{1. A novel dual-branch speech enhancement network is proposed, which integrates the complementary strengths of both ANN and SNN architectures to achieve superior performance.}
\item{2. The ANN branch introduces BandSplit and TF-Mamba blocks to model speech characteristics with low computational complexity. In parallel, the SNN branch proposes SFEG and ITB to mitigate inherent information loss.}
\item{3. The interaction and TF-CAF modules are designed to achieve branch information fusion, optimally adapted to the characteristics of speech signals.}
\item{4. DBHN-Net achieves state-of-the-art performance on three public datasets while reducing computational complexity by an average of 7.5 times compared to baseline models.}
\end{list}

\section{Related Work}
As shown in Table \ref{table1}, we summarize the structural characteristics of various speech enhancement networks, as well as their performance in terms of computational complexity and energy consumption, allowing for a more intuitive understanding of the shortcomings of ANN-based and SNN-based speech enhancement networks.

\subsection{ANN-Based Speech Enhancement Model}
In recent years, ANN-based speech SE models have demonstrated remarkable performance \cite{ai2,tpami3}. Representative models like CRN \cite{CRN}, GCRN \cite{GCRN}, and DPRNN \cite{DPRNN} process complex spectral inputs in the time-frequency domain, effectively incorporating phase information to enhance performance \cite{ai3}. As ANN architectures evolve, SE models have become increasingly tailored to speech characteristics. FullSubNet \cite{fullsubnet} pioneered the full-band and sub-band parallel processing approach to improve the capability to model low-frequency speech components, with its enhanced version FullSubNet+ \cite{fullsubnet+} further improving cross-band modeling capabilities. To address the ``compensation effect'' between magnitude and phase, CTS-Net \cite{ctsnet} introduced a two-stage paradigm that derives complementary phase information from magnitude spectra. GaG-Net \cite{gagnet} advanced this architecture through parallel dual-path processing for more complete information separation. TaylorNet \cite{li2022taylor} employs Taylor series expansion (zero-order and higher-order terms) for feature representation, coupled with multi-stage distributed processing for enhanced performance. For cross-domain information integration, CompNet \cite{compnet} combines time-domain and T-F domain features through complementary cross-domain optimization \cite{bsdb}. Various sequence modeling modules have also been incorporated - DBT-Net \cite{DBTNET} developed a Transformer-based dual-path framework that achieves excellent results despite high complexity, while BSDB-Net \cite{bsdb} combines band-split processing with Mamba blocks to maintain high performance with significantly reduced computational load.
\subsection{Spiking Neural Networks}
Recently, Spiking Neural Networks (SNNs) have attracted significant attention due to their brain-inspired mechanisms that substantially reduce energy consumption \cite{yudi,snnjia}. SNNs have been widely applied in computer vision, primarily implemented through ANN-SNN conversion strategies and surrogate gradient approaches. In fundamental SNN research, \cite{1111} proposed similarity-sensitive contrastive learning to maximize mutual information between ANN and SNN intermediate representations, SpikingBERT \cite{SpikingBERT} introduced a novel ANN-SNN based knowledge distillation method, TC-LIF \cite{TC-LIF} developed Two-Compartment (TC) -LIF neurons to enhance long-sequence modeling. For applied SNN research, ESDNet \cite{ESDNet} integrated SNN features with image deraining, and SpikeLM \cite{SpikeLM} proposed an elastic bi-spiking mechanism for both discriminative and generative tasks. However, current applications of SNNs in speech enhancement remain limited: DPSNN \cite{dpsnn} adapted DPRNN into an SNN-based architecture for speech recognition, TD-STNet \cite{TD-STNet} designed a temporally dynamic spiking Transformer network for speech enhancement, and Spiking-UNet \cite{u-netspiking} modified U-Net into an SNN-based speech enhancement system. The Intel Neuromorphic DNS Challenge 2023 specifically focused on SNN applications, where Spiking-FullSubNet \cite{Spiking-FullSubNet} won first place by adapting FullSubNet into an SNN-based speech enhancement model.

\begin{figure*}[!t]
\begin{center}
\centerline{\includegraphics[width=1\textwidth]{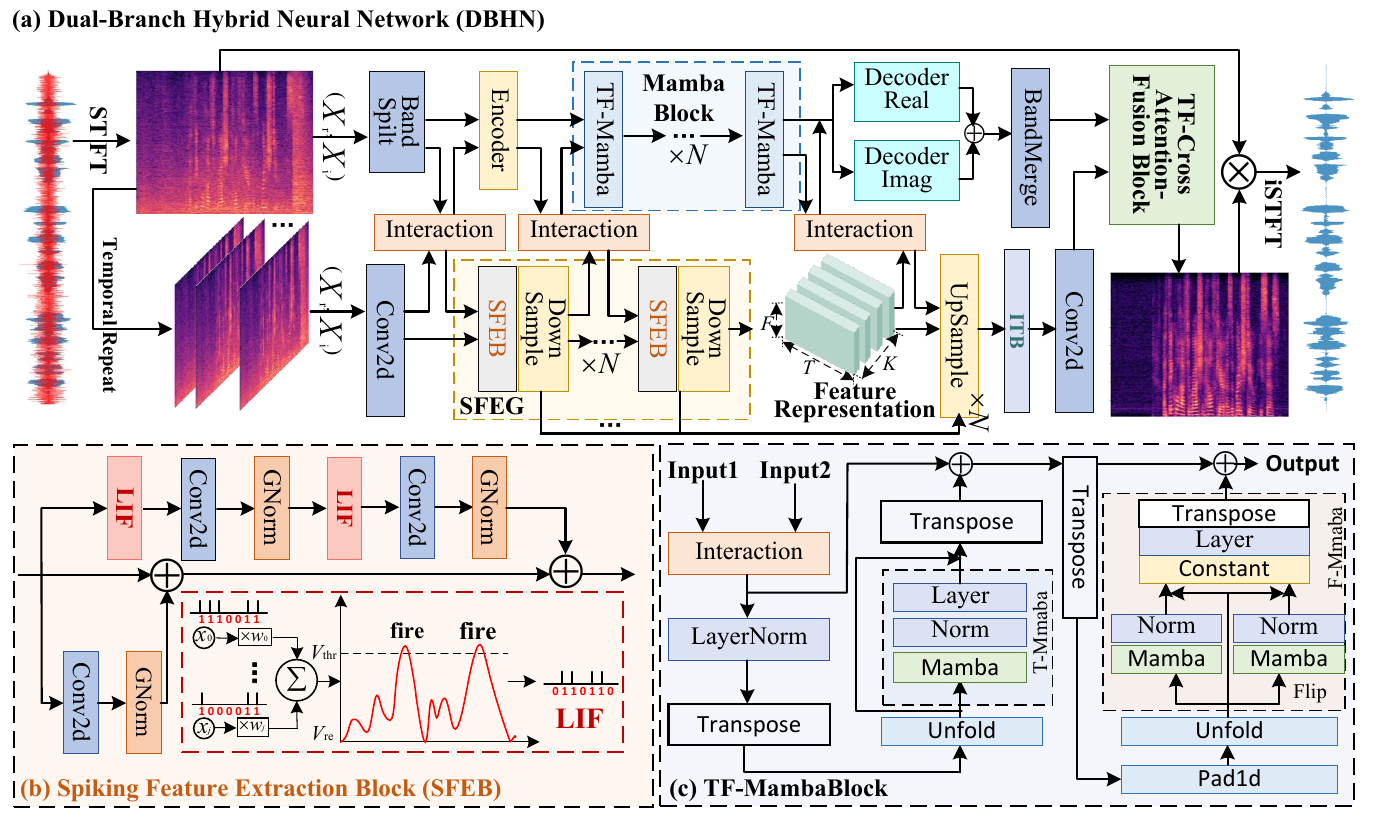}}
\caption{\rev{(a) The proposed Dual-Branch Hybrid Neural Network takes complex spectra as input. The upper branch is the ANN pathway, primarily consisting of a BandSplit module and a TF-Mamba sequential modeling module. The lower branch represents the SNN pathway, mainly comprising a Spiking Feature Extraction Group and an Information Transformation Block. (b) The proposed Spiking Feature Extraction Block primarily employs LIF units to convert information into binary (0/1) spike signals. (c) The proposed TF-Mamba Block mainly contains T-Mamba and F-Mamba components for sequential feature modeling.}}
\label{fig1}
\end{center}
\end{figure*}

\begin{figure}[!t]
\begin{center}
\centerline{\includegraphics[width=0.5\textwidth]{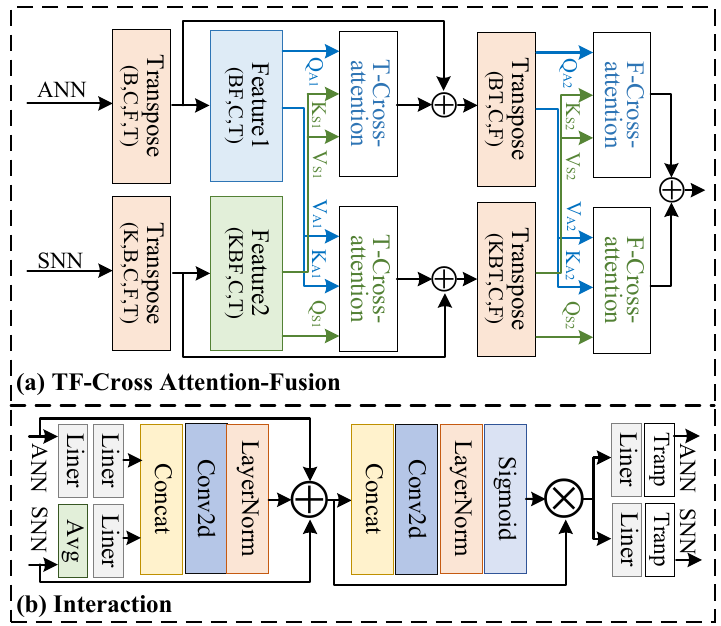}}
\caption{(a) The proposed TF-Cross Attention Fusion module integrates dual-branch information through T-Cross Attention (time-domain) and F-Cross Attention (frequency-domain) operations. (b) The designed Interaction module facilitates cross-branch information exchange throughout all network stages.}
\label{fig2}
\end{center}
\end{figure}

\section{The Proposed Model}
The received noisy mixture signal can be represented in the Short-Time Fourier Transform (STFT) domain as:
\begin{equation}
\label{q1}
Y_{(t, f)} = X_{(t, f)} + N_{(t, f)},
\end{equation}
where $\left\{Y, X, N\right\}\in\mathbb{C}^{T\times F}$ denote the mixture, clean and noise signals, respectively. $t\in \{1,\cdots, T\}$ denotes the frame index, and $f\in \{1,\cdots,F\}$ is the frequency index. Here, T corresponds to the total count of time frames, and F represents the number of frequency bins per spectrum.
\subsection{Data Preprocessing}
The noisy speech input is first transformed into a complex-valued spectrum through the STFT. Following the approach in \cite{li2022taylor}, its real and imaginary (RI) components are stacked along the channel axis, denoted as $\mathrm{Y_1\in\mathbb{R}^{2\times T\times F}}$. The ANN-based branch directly processes this complex spectrum as input. However, due to the unique spatiotemporal characteristics of Spiking Neural Networks (SNNs), we first encode the input complex spectrum $Y$ along the first dimension to generate a sequence $\mathrm{\overline{Y}={\{Y_{k}\}}^{K}_{k=1}\in\mathbb{R}^{K\times 2\times T\times F}}$. Specifically, we replicate each single-frame complex spectrum $\mathrm{Y_{k}}$ as input for every timestep k{\footnote{Due to the inherent sequential nature of speech, there exists a time or frame dimension in the speech processing field. In a Spiking Neural Network (SNN), however, there is also a repeated spiking dimension due to the network operation step. To avoid the confusion between the two conceptions, unless otherwise specified, the repeated spiking dimension is denoted as k and the time axis of the speech is notated as t.}}.
\subsection{Dual-Branch Network Architecture}
The proposed network is shown in Figure \ref{fig1}(a). The proposed DBHN primarily consists of an ANN branch, an SNN branch, and a final TF-Cross Attention Fusion Block (TF-CFB) . Time-domain speech signals are transformed into time-frequency representations via STFT, with the ANN branch directly taking the complex-valued spectra as input. Due to the SNN's unique iterative processing across timesteps, the complex spectra are first replicated $K$ times to form a new $(K, B, C, F, T)$ dimensional input for the SNN branch.

The complex spectrum in the ANN branch is first processed by the Band-Split module, which sequentially divides the spectrum along the frequency dimension from low to high frequencies. The iteratively processed data in the SNN branch is initially fed into a 2D convolution for feature extraction. The information from both branches is then input into the first Interaction module for cross-branch exchange. Subsequently, the ANN branch combines the output of the first Interaction module and the Band-Split module as input to the Encoder, which performs feature extraction on the ANN branch data. Meanwhile, the SNN branch takes the output of the first Interaction module and the 2D convolution as input to the first Spiking Feature Extraction Block (SFEB) and downsampling module for feature modeling. The data from both branches then enters the second Interaction module for further information exchange. Next, the ANN branch feeds the output of the second Interaction module and the Encoder into the Mamba block, which consists of $N$ TF-Mamba modules for sequential modeling. The SNN branch processes the output of the second Interaction module and the first SFEB through subsequent $N$ SFEBs and downsampling modules for sequential feature extraction. The data then proceeds to the third Interaction module for another round of information fusion. Finally, the ANN branch passes the output of the third Interaction module and the mamba block into the decoder, followed by the Band-Merge module to reconstruct the originally split complex spectrum, yielding the final ANN branch output. The SNN branch processes the output of the third Interaction module and the last downsampling module through $N$ upsampling modules for data reconstruction, then refines the features via the Information Transformation Block before passing them through a final 2D convolution to produce the SNN branch's output.

To further integrate the dual-branch information, the outputs from both branches are fed into the TF-CAF Block for time-domain and frequency-domain fusion, generating enhanced complex spectra that are finally transformed into enhanced speech via ISTFT.

\subsection{ANN Branch Architecture}
\begin{figure}[!t]
\begin{center}
\centerline{\includegraphics[width=0.5\textwidth]{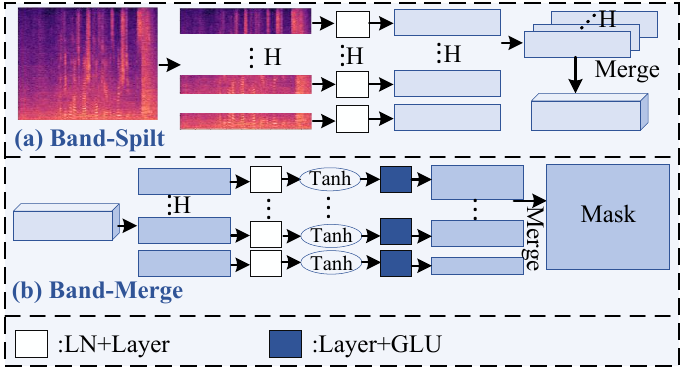}}
\caption{\rev{(a) The Band-Split module, operating at the initial stage of the ANN branch, decomposes the complex spectrum along the frequency dimension. (b) The Band-Merge module, functioning at the final stage of the ANN branch, reconstructs the segmented complex spectra.}}
\label{fig3}
\end{center}
\end{figure}

\begin{figure}[!t]
\begin{center}
\centerline{\includegraphics[width=0.5\textwidth]{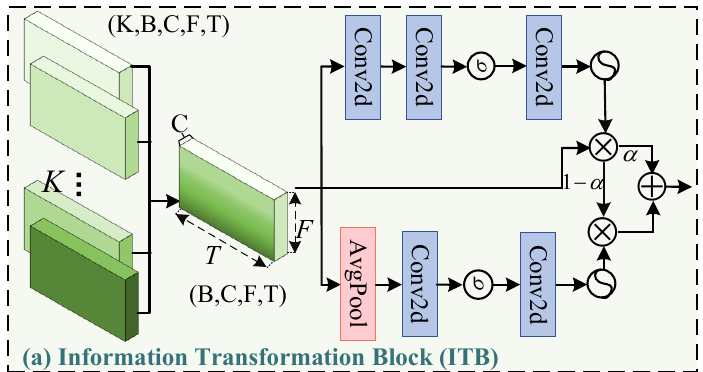}}
\caption{The proposed Information Transformation Block operates at the terminal stage of the SNN branch to further refine information processing.}
\label{fig4}
\end{center}
\end{figure}

\subsubsection{Band-Split}
As shown in Figure \ref{fig3}(a), the input noisy complex spectrum $X$ is partitioned \rev{along the frequency dimension into a series of nonoverlapping frequency bands $\{B\}^H_{i=1}$, with each subband expanded into an embedding $N-dimensional$. The $H$ bands are then aggregated into a new tensor $F$, which can be formally expressed as}:
\begin{equation}
\label{q2}
\mathrm{B_{i}\in \left \{ B_{1}, B_{2},..., B_{H} \right \}, B_{i}\in \mathbb{R}^{T\times F_{i}}} ,
\end{equation}
\begin{equation}
\label{q3}
\mathrm{C_{i}=FC_1(LN_1(B_{i})), C_{i}\in \mathbb{R}^{T\times N}},
\end{equation}
\begin{equation}
\label{q4}
\mathrm{X_{ANN}^1 = Concat_1\left(C_{1}, C_{2},..., C_{H}\right)\in \mathbb{R}^{H\times T\times N}},
\end{equation}
where $\mathrm{B_{i}}$ represents the i-th partitioned frequency subband, $\mathrm{\{FC_1, LN_1\}}$ denote the fully-connected layer and layer normalization (LN) respectively, $\mathrm{Concat_1\left(\cdot\right)}$ indicates the concatenation operation, and $\mathrm{X_{ANN}^1}$ corresponds to the output of the Band-Split module.

\subsubsection{Encoder}
The Encoder module takes the outputs from both the Band-Split module and the first Interaction module as inputs. First, it concatenates these two inputs along the second dimension, then processes them through a 2D convolution, LayerNorm, and Sigmoid activation to generate a mask. This mask is multiplied with the Interaction module's output and added to the Band-Split module's output. The fused data subsequently passes through another 2D convolution, LayerNorm, and ReLU activation to produce the final output, which can be mathematically represented as:
\begin{equation}
\label{q5}
\mathrm{D=Concat(X_{ANN}^1,Inter_1)},
\end{equation}
\begin{equation}
\label{q6}
\mathrm{D_{Mask}=Sigmoid(LN(Conv2d(D)))},
\end{equation}
\begin{equation}
\label{q7}
\mathrm{D_2=X_{ANN}^1+(Inter_1 \otimes D_{Mask})},
\end{equation}
\begin{equation}
\label{q8}
\mathrm{X_{ANN}^2=Relu(LN(Conv2d(D_2)))},
\end{equation}
where $\mathrm{X_{ANN}^1, Inter_1}$ represent the output of Band-Split module and the output of the first Interaction module, $\mathrm{X_{ANN}^2}$ denotes the output of the Encoder, $\otimes$ denotes the elementary multiplication operation, $\mathrm{\{Sigmoid, Relu \}}$ denote the Sigmoid activation function and the Relu activation function.

\subsubsection{TF-Mamba block}
As shown in Figure \ref{fig1}(c), the core of the Mamba Block lies in its use of Mamba for sequence modeling. Mamba is a selective state space model (SSM) that enhances traditional SSMs by introducing input-dependent processing , hardware-aware algorithms, and superior sequence modeling capabilities. Its discrete state computation can be expressed as:
\begin{equation}
\label{q9}
h_{t}=\overline{\mathbf{A}} h_{t-1}+\overline{\mathbf{B}} x_{t},
\end{equation}
\begin{equation}
\label{q10}
y_{t}=\mathbf{C} h_{t},
\end{equation}

Here, $\overline{\mathbf{A}}$ represents the discretized state matrix, both $\overline{\mathbf{B}}$ and $\mathbf{C}$ denote the discretized projection matrices, and $h_{t}$ corresponds to the hidden state. Through the discretization process, Mamba converts continuous parameters into discrete ones, enabling effective processing of discrete data.

The Mamba block comprises multiple TF-Mamba modules, where the first TF-Mamba integrates the outputs from both the Encoder and the second Interaction module, \rev{the second TF-Mamba module takes the output of the first TF-Mamba and the output of the second Interaction module as its input. Here, $N$ is set to 4. Admittedly, increasing the value of $N$ could further improve performance, but it would also significantly increase the computational complexity. To balance the relationship between the two, we ultimately determined that $N$ = 4 and the hidden layer parameter of TF-Mamba as 128 can achieve good results with low computational complexity.} After fusing these inputs, it reshapes the data from dimensions $(B, F, N, T)$ to $(BF, N, T)$ for temporal modeling in the T-Mamba module. This module contains a causal (unidirectional) Mamba layer, LayerNorm, and a linear projection layer, with the unidirectional architecture ensuring strict causality. The mathematical representation is:
\begin{equation}
\label{q11}
\mathrm{E=Concat (X_{ANN}^2,Inter_2)},
\end{equation}
\begin{equation}
\label{q12}
\mathrm{E_{Mask}=Sigmoid(LN(Conv2d(E)))},
\end{equation}
\begin{equation}
\label{q13}
\mathrm{E_2=X_{ANN}^2+(Inter_2 \otimes D_{Mask})},
\end{equation}
\begin{equation}
\label{q14}
\mathrm{E_3=T-Mamba(Unfold(Trans(LN(E_2))))},
\end{equation}
where $\mathrm{X_{ANN}^2 and Inter_2}$ represent the output of Encoder module and the output of the second  Interaction module, respectively $\mathrm{E_3}$ represents the intermediate features processed by the T-Mamba module within TF-Mamb, $\mathrm{\{T-Mamba( \cdot), Trans(\cdot)\}}$ represent unidirectional Mamba-based modeling of the data and dimensional transformation of the data, respectively.

Subsequently, the output of T-Mamba undergoes a residual connection before being reshaped from dimensions $\mathrm{(BF, N, T)}$ to $\mathrm{(BT, N, F)}$ for processing in F-Mamba, which performs frequency-domain modeling. F-Mamba shares a similar structure with T-Mamba, with the key distinction being its use of bidirectional Mamba processing. As illustrated in Figure \ref{fig1}(c), this is implemented by flipping the input data and processing it through dual parallel Mamba paths. The mathematical representation is as follows:
\begin{equation}
\label{q15}
\mathrm{E_4=Unfold(Trans((E_3+E)))}.
\end{equation}
\begin{equation}
\label{q16}
\mathrm{X_{ANN}^3=F-Mamba(E_4)},
\end{equation}
where $\mathrm{X_{ANN}^3}$ represents the output after sequence modeling through the Mamba Block (containing N TF-Mamba modules), and $\mathrm{F-Mamba(\cdot)}$ represents the F-Mamba module illustrated in Figure \ref{fig1}(c).
\subsubsection{Decoder}
After sequential modeling through N TF-Mamba modules, the outputs from both the Mamba block and the third Interaction module are fed into the Decoder for feature reconstruction. The decoding process specifically consists of a 2D transposed convolution, LayerNorm, ReLU activation, and 2D convolution. The Decoder processes the real and imaginary components separately through identical architectures. Taking one component as an example:
\begin{equation}
\label{q17}
\mathrm{F=Concat (X_{ANN}^3,Inter_3)},
\end{equation}
\begin{equation}
\label{q18}
\mathrm{F_{Mask}=Sigmoid(LN(Conv2d(F)))},
\end{equation}
\begin{equation}
\label{q19}
\mathrm{F_2=X_{ANN}^3+(Inter_3 \otimes F_{Mask})},
\end{equation}
\begin{equation}
\label{20}
\mathrm{X_{ANN}^4= Conv2d(Relu(LN(ConvTrans(F_2))))},
\end{equation}
where $\mathrm{X_{ANN}^4}$ denotes the output of the Decoder module, $\mathrm{ConvTrans(\cdot)}$ represents the 2D transposed convolution operation.

\subsubsection{Band-Merge}

At the final stage of the ANN branch, the data undergoes frequency-band reconstruction through the Band-Merge layer to restore the original complex spectrum. As shown in Figure \ref{fig3}(b), \rev{the module takes input $\mathrm{G\in \mathbb{R}^{H\times T\times N}}$, splits it along the frequency dimension into subbands $\mathrm{G_{i}\in \mathbb{R}^{T\times N}}$ (where $i \in \{1,\dots,H\}$),} then processes each through a linear layer, LayerNorm, Tanh activation, and GLU layer to recover the original subband dimensions before finally merging them into the complete complex spectrum, as mathematically represented below:

\begin{equation}
\label{21}
\mathrm{M_i= GLU(Tanh(FC(LN(G_i)))},
\end{equation}
\begin{equation}
\label{22}
\mathrm{X_{ANN}^{fianl}= Merge(G_1,...,G_H),X_{ANN}^{fianl} \in \mathbb{R}^{F\times T\times N}},
\end{equation}
where $\mathrm{M_i}$ denotes the i-th band feature, $\mathrm{\{Tanh,GLU\}}$ denote the Tanh activation function and gated linear unit, respectively. $\mathrm{Merge (\cdot)}$ is the concatenation operation along the frequency axis.

\subsection{SNN Branch Architecture}

\subsubsection{Spiking Feature Extraction Block}
The proposed architecture is illustrated in Figure \ref{fig1}(b). The preprocessed complex spectral sequence is first fed into a Conv2d, after which the information is directed to both the first Interaction module and the first SFEB. The first SFEB takes inputs from the Conv2d layer and the first Interaction module, and processes them through N SFEB+DownSample structures for feature modeling. Without loss of generality, we use the first SFEB as an example, where the input features are divided into three branches for processing.

The first branch converts the original signal into 0/1 spike signals via a Leaky Integrate-and-Fire (LIF) node \cite{lif, lifxin}, followed by feature extraction using a Conv2d layer with a 3×3 kernel. The processed signal then passes through a Group Normalization (GN) layer to accelerate training convergence. These operations are repeated to generate the first branch's output. The second branch extracts features through Conv2d and GN layers. Finally, the outputs from both branches are fused with the original input to produce the final output, which is given by:
\begin{equation}
\label{23}
\mathrm{X_{SNN}^1= Conv2d(\overline{Y_k})},
\end{equation}
\begin{equation}
\begin{split}
\mathrm{A_k^1 = GN(Conv2D(LIF(X_{SNN}^1, Inter_1)))},
\end{split}
\label{24}
\end{equation}
\begin{equation}
\label{25}
\mathrm{A_k^2 = GN(Conv2D(X_{SNN}^1, Inter_1))},
\end{equation}
\begin{equation}
\label{26}
\mathrm{X_{SNN}^2 = A_k^1 \oplus A_k^2 \oplus Intercation(X_{SNN}^1, Inter_1)},
\end{equation}
where $\mathrm{\overline{Y_k}}$ represents the input at the k-th timestep. $\oplus$ denotes the elementary sum operation. $\mathrm{\{X_{SNN}^1, X_{SNN}^2\}}$ represent the outputs of the 2D convolutional module and the SFEB module, respectively. $\mathrm{Interaction(\cdot)}$ refers to the module denoted by subsection F of Section III.

In this work, we adopt the Leaky Integrate-and-Fire (LIF) neuron model for its efficiency and mathematical convenience. Its core component is the membrane potential, an internal state that decays with a time constant $\lambda$. A spike is emitted once this potential surpasses a set threshold, triggering a reset that returns it to a base level. This entire dynamic is captured by the subsequent discrete-time formulation:
\begin{equation}
\boldsymbol{H}_{k}^{n}=\boldsymbol{U}_{k-1}^{n}+\frac{1}{\tau}\left(\boldsymbol{X}_{k}^{n}-\left(\boldsymbol{U}_{k-1}^{n}-V_{\text {reset }}\right)\right),
\end{equation}
\begin{equation}
\boldsymbol{S}_{k}^{n}=\Theta\left(\boldsymbol{H}_{k}^{n}-V_{\text {thr }}\right)=\left\{\begin{array}{ll}
1 & \boldsymbol{{H}}_{k}^{n}>=V_{\text {thr}} \\
0 & \text { otherwise },
\end{array}\right. \\
\end{equation}
\begin{equation}
\boldsymbol{U}_{k}^{n}=\left(\beta \boldsymbol{H}_{k}^{n}\right) \odot\left(\mathbf{1}-\boldsymbol{S}_{k}^{n}\right)+V_{\text {reset }} \boldsymbol{S}_{k}^{n},
\end{equation}
where $k$ and $n$ denote the $k$-th  spiking time step and the $n$-th layer, respectively. The term $\boldsymbol{H}_{k}^{n}$ signifies the membrane potential, produced by integrating the temporal input $\boldsymbol{U}_{k-1}^{n}$ with the spatial input $\boldsymbol{X}_{k}^{n}$ . The membrane time constant is given by $\tau$, while $\beta$ is the decay factor. The activation threshold $V_{thr}$ controls the output: if exceeded, a spike $\boldsymbol{S}$ is generated; otherwise, the output remains zero, as defined by the step function $\Theta(\cdot)$. Following a spike, the potential is reset to $V_{reset}$ .
\subsubsection{DownSampling Block }
After processing through the SFEB, the data is fed into the DownSample module for feature extraction, which consists of an LIF Node, Conv2d layer, and GN layer:
\begin{equation}
\label{30}
\mathrm{X_{SNN}^3 = GN(Conv2D(LIF(X_{SNN}^2)))},
\end{equation}
where $X_{SNN}^3$ represents the output of the DownSample module.
\subsubsection{UpSampling Block}
After N SFEB+DownSample operations, an intermediate representation is formed, which is then restored to the original dimensions through $N$ SFEB+UpSample blocks—their structure matching the earlier description. The output after these $N$ SFEB+UpSample blocks is denoted as $X_{SNN}^4$. Notably, the input to the first SFEB+UpSample block also includes the output from the third Interaction module.
\subsubsection{Information Transformation Block}
Figure \ref{fig4} shows the proposed ITB structure. To further refine features and capture detailed information, we need to convert discrete values into continuous ones. For this purpose, we designed a feature refinement module that transforms discrete spike values into continuous information representations while extracting multi-perspective features to minimize information loss. The module processes information through two parallel branches:
\begin{equation}
\label{31}
\mathrm{Q_1=Rule(Conv2D(Sig(Conv2D(X_{SNN}^3))))},
\end{equation}
\begin{equation}
\label{32}
\mathrm{Q_2=Rule(Conv2D(Sig(AvgPool(X_{SNN}^3))))},
\end{equation}
\begin{equation}
\label{33}
\mathrm{X_{SNN}^{final} = (Q_1 \otimes X_{SNN}^3)\oplus(Q_2 \otimes (1-Q_1))},
\end{equation}
\begin{equation}
\label{34}
\mathrm{X_{SNN}^{final} = (Q_1 \otimes X_{SNN}^3)\oplus(Q_2 \otimes (1-Q_1))},
\end{equation}
where $\mathrm{\{X_{SNN}^3, Q_1 ,Q_2, X_{SNN}^{final}\}}$ represent the input of this module, the output of the first branch, the output of the second branch, and the final output of the SNN branch, respectively.
\subsection{TF-Cross Attention Fusion Block}
The TF-Cross Attention Fusion module integrates the final outputs from the ANN and SNN branches, performing a dual-domain fusion that operates simultaneously in the time and time-frequency domains, as illustrated in Figure \ref{fig2}(a). Specifically, the data from each branch is first transformed to the same dimension, then processed by a multi-head attention module. The mathematical formulation is expressed as follows:
\begin{equation}
\label{35}
\mathrm{F_{ANN}^{T} =TCA(Re(X_{ANN}^{final}),Re(X_{SNN}^{final}))+ X_{ANN}^{final}},
\end{equation}
\begin{equation}
\label{36}
\mathrm{F_{SNN}^{T} =TCA(Re(X_{SNN}^{final}),Re(X_{ANN}^{final}))+ X_{SNN}^{final}},
\end{equation}
\begin{equation}
\label{37}
\begin{split}
\mathrm{F_{final}} =\mathrm{FCA}(\mathrm{Re(F_{ANN}^{T}}),\mathrm{Re(F_{SNN}^{T})})\\
          + \mathrm{FCA}(\mathrm{Re(F_{ANN}^{T}),Re(F_{SNN}^{T})}),
\end{split}
\end{equation}
where $\mathrm{\{F_{ANN}^{T}, F_{SNN}^{T}, F_{final} \}}$ represent the ANN branch data processed by T-Cross, the SNN branch data processed by T-Cross, and the final output data of the TF-CAF module, respectively. $\mathrm{\{Re(\cdot) \}}$ denotes the reshaping of tensor dimensions. $\mathrm{\{TCA(\cdot), FCA(\cdot) \}}$ represent multi-head attention structures, where both modules share identical architectures but process different input data.
\subsection{Interaction Block}
Throughout the proposed network architecture, we employ multiple Interaction modules to enhance information exchange between the two branches. As illustrated in Figure \ref{fig2}(b), each module takes inputs from both branches and delivers outputs back to them. The mathematical formulation is expressed as follows:
\begin{equation}
\label{38}
\begin{split}
\mathrm{OUT_{1}} = \mathrm{LN(Conv2d(Concat(Liner(ANN_{in})},\\
                    \mathrm{Avg(SNN_{in}))))+ANN_{in}+SNN_{in}},
\end{split}
\end{equation}
\begin{equation}
\label{39}
\begin{split}
\mathrm{OUT_{2}} = \mathrm{Sig(LayerNorm(Conv2d(OUT_{1})))}\\
          \otimes \mathrm{OUT_{1}},
\end{split}
\end{equation}
\begin{equation}
\label{40}
\begin{split}
\mathrm{ANN_{out} = Liner(Transpose(OUT_{2}))},\\
\mathrm{SNN_{out} = Liner(Transpose(OUT_{2}))},
\end{split}
\end{equation}
where $\mathrm{\{ANN_{in}, SNN_{in}, ANN_{out}, SNN_{out} \}}$ denote the input of the ANN branch, input of the SNN branch, output of the ANN branch, and output of the SNN branch, respectively. $\mathrm{\{Transpose(\cdot)\}}$ denotes tensor transposition.

\section{Experimental Setup}
\subsection{Datasets}
We employed three public datasets and conducted comparisons against baseline models.
\subsubsection{WSJ0-SI84+DNS-Challenge}
We utilized the WSJ0-SI84 corpus, which comprises 7,138 clean utterances from 83 speakers (42 male, 41 female) \cite{wsj0}. The training and validation sets were constructed from 77 speakers, containing 5,428 and 957 utterances, respectively. For testing, 150 utterances were randomly drawn from the remaining 6 unseen speakers. Noisy-clean training pairs were synthesized by approximately 20,000 noise samples were randomly selected from the noise library of DNS-Challenge \cite{DNSChallenge}, which were concatenated together for a total duration of 55 hours. During training and validation set construction, the SNR range was set to [-5, 0] dB with 1 dB increments. For each synthesis, a random utterance from the training set was mixed with a noise segment of equal length, sampled at the target SNR. This process was repeated 150,000 times for training and 10,000 times for validation, yielding roughly 300 hours of data. The test set includes two highly non-stationary noise types from the NOISEX92 \cite{NOISEX-92} database: babble and factory1.

\subsubsection{VoiceBank+Demand}
The VoiceBank+Demand \cite{voicebank} dataset comprises recordings from 30 speakers. Following the common practice established in prior work \cite{usevoicebank}, 28 speakers were allocated for training and the remaining two for testing. This setup provides a training set of 11,572 noisy-clean utterance pairs, mixing 10 noise types (8 from the Demand noise database and 2 artificial noises) at 4 SNR levels: 0 dB, 5 dB, 10 dB, and 15 dB. The test set includes 824 pairs with 5 unseen noise types from Demand, evaluated at SNRs of 2.5 dB, 7.5 dB, 12.5 dB, and 17.5 dB.
\subsubsection{DNS-Challenge}
The DNS-Challenge dataset is a large-scale collection where clean speech is sourced from the LibriSpeech corpus \cite{DNSChallenge}, while noise clips come from AudioSet, Freesound, and the DEMAND corpus. We selected 40,000 high-quality utterances (approximately 89 hours) from DNS4, then synthesized 240,000 training samples (about 530 hours) with SNRs ranging from -5 to 0 dB by mixing these clean utterances with 50,000 noise clips from DNS-Challenge, along with a 10,000-sample validation set. Evaluation was conducted on the official test set containing 150 synthetic noisy samples without reverberation, with SNRs randomly distributed between 0 and 20 dB, consistent with other compared models.

\subsection{Implementation Setup}
The training set underwent the following preprocessing pipeline: all speech signals were standardized at a 16 kHz sampling rate, and each utterance was adjusted to a fixed 8-second duration through truncation or zero-padding to stabilize training. Signal framing employed a 20 ms Hanning window with a 50\% overlap. A 320-point FFT was then applied to these frames to generate time-frequency representations with 161 frequency bins. Following the methodology in \cite{YASUOYOUXIAO1}, we applied a power-law compression (exponent 0.5) to the magnitude spectra of both input and target (i.e., $|\textbf{X}|^{0.5}$, $|\textbf{S}|^{0.5}$).

Our implementation used PyTorch 1.6.0 and the Adam optimizer. Training was configured with an initial learning rate of 5e-4, which was reduced by half if the validation loss plateaued for two consecutive epochs. The training process was terminated after three epochs showed no improvement and was run for a maximum of 60 epochs with a batch size of 3.

\subsection{Baseline Models}
We benchmarked our proposed model against 11 established methods on the WSJ0-SI84 dataset. ConvTasNet \cite{Conv-tasnet} and DPRNN \cite{DPRNN} are time-domain speech enhancement models, while LSTM \cite{LSTM}, CRN \cite{CRN}, GCRN \cite{GCRN}, DCCRN \cite{dccrn}, FullSubNet \cite{fullsubnet}, CTSNet \cite{ctsnet}, and GaG-Net \cite{gagnet} operate in the time-frequency domain. BSDBNet \cite{bsdb} introduces Mamba for dual-branch sequence modeling, and spiking-UNet adapts the U-Net architecture into an SNN framework. GCRN and DCCRN extend CRN, FullSubNet, which integrates full-band and sub-band processing; CTSNet and GaG-Net, which employ parallel and serial two-stage frameworks for decoupling amplitude and phase, respectively. BSDBNet \cite{bsdb} is particularly noted for its effective balance of model efficiency and performance.
\begin{table}[]
\begin{center}
\caption{Ablation studies on the proposed dual-branch architecture were conducted by separately disabling the ANN branch and SNN branch to evaluate their individual contributions to model performance.}
\begin{tabular}{c|ccc}
\toprule
\textbf{Model}          & \textbf{PESQ}          & \textbf{ESTOI (in\%)}          & \textbf{SI-SDR (in dB)}         \\ \midrule
\textbf{DBHN-Net (OURS)} & \textbf{3.17} & \textbf{83.46} & \textbf{12.16} \\
w/o ann-branch & 2.75          & 74.91          & 10.41           \\
w/o snn-branch & 2.81          & 77.32          & 10.89          \\ \bottomrule
\end{tabular}
\label{table2}
\end{center}
\end{table}
On the VoiceBank+Demand dataset, 17 baseline methods were selected for comparison with the proposed model. SEGAN \cite{segan}, MMSEGAN \cite{mmsegan}, and MetricGAN \cite{metricgan} are generative speech enhancement models, while Wavenet \cite{van2016wavenet} operates in the time domain. SRTNET \cite{srtnet} achieves time-domain speech enhancement through stochastic refinement. PHASEN \cite{Phasen}, MHSASPK \cite{MHSA-SPK}, DCCRN, TSTNN \cite{tstnn}, S4NDUNet \cite{s4sd}, FDFnet \cite{FDFnet}, CSTnet, and GaG-net operate in the time-frequency domain. CompNet \cite{compnet} spans both time and time-frequency domains. BSDBNet incorporates Mamba into a dual-branch network to balance computational complexity and performance. U-Net-Spiking and FullSubNet-Spiking are two SNN-based speech enhancement networks that modify activation functions to convert original ANN architectures into SNN frameworks.

On the DNS-Challenge dataset, we selected 14 baseline models for comparison with DBHN-Net. First, NSNet \cite{DNSChallenge} is the baseline model provided in DNS-Challenge 2020, which combines GRU modeling and loss optimization based on speech distortion control; DTLN \cite{dtln} is based on dual-transform domain learning, including filtering operations in the STFT domain and the latent domain, with each transform domain using LSTM as the modeling unit; DCCRN, FullSubNet, CTS-Net, and GaG-Net are the same as previously described; DCCRN+ is similar to the DCCRN model structure and replaces the real-valued LSTM with a complex-valued LSTM; TaylorSENet \cite{li2022taylor} combines Taylor expansion with the refinement theory of speech enhancement; IIFC-Net \cite{IIFC-Net} is a single-channel speech enhancement network using High-Order Information Interaction and Feature Calibration; MFNet \cite{MFNet} is a direct and simple network that can map both speech and inverse noise; SICRN \cite{SICRN} combines State space model and Inplace Convolution; CMGAN-Bispectra \cite{CMGANB} incorporates bispectral features in place of the original DDN encoder module, enabling more effective modeling of frequency correlations and phase; In contrast, TSDT-Net \cite{TSDT-Net} employs a dual-stage, ultra-low complexity architecture designed to deliver superior denoising performance under strict constraints on parameters and computational budget.

\setlength{\tabcolsep}{2.6pt}  
\begin{table}[]
\begin{center}
\caption{Ablation experiments on the Mamba sequence modeling module in the ANN branch. {LSTM, Transformer} means replacing the sequence modeling Mamba in the middle with LSTM or Transformer for the experiments.}
\begin{tabular}{c|cccc}
\toprule
\textbf{Model}          & \textbf{PESQ} & \textbf{ESTOI (in\%)} & \textbf{SI-SDR (in dB)} & \textbf{MACs}    \\ \midrule
\textbf{DBHN-Net (OURS)} & \textbf{3.17} & \textbf{83.46} & \textbf{12.16}  & \textbf{1.32G/s} \\
LSTM                    & 3.01          & 79.09          & 11.18           & 8.68G/s          \\
Transformer             & 3.13          & 82.12          & 11.97           & 22.05G/s         \\ \bottomrule
\end{tabular}
\label{table3}
\end{center}
\end{table}

\begin{table}[]
\begin{center}
\caption{Ablation experiments on the SFEB and ITB modules in the SNN branch. ``w/o SFEB'' and ``w/o ITB'' refer to experiments where the SFEB module is directly replaced with a simple LIF unit and the ITB module is directly removed, respectively.}
\begin{tabular}{c|ccc}
\toprule
\textbf{Modle}          & \textbf{PESQ} & \textbf{ESTOI (in\%)} & \textbf{SI-SDR (in dB)} \\ \midrule
\textbf{DBHN-Net (OURS)} & \textbf{3.17}          & \textbf{83.46}          &\textbf{12.16}           \\
w/o SFEB       & 3.07          & 81.82          & 11.33           \\
w/o ITB        & 3.11          & 82.04          & 11.92           \\ \bottomrule
\end{tabular}
\label{table4}
\end{center}
\end{table}

\begin{table}[]
\begin{center}
\caption{Ablation studies on the proposed TF-CAF and Interaction modules were conducted, specifically examining three configurations: (1) removal of the TF-CAF module alone, (2) removal of the Interaction module alone, and (3) removal of both modules.}
\begin{tabular}{c|ccc}
\toprule
\textbf{Model}                                         & \textbf{PESQ}          & \textbf{ESTOI (in\%)}          & \textbf{SI-SDR (in dB)}         \\ \midrule
\multicolumn{1}{c|}{\textbf{DBHN-Net (OURS)}}           & \textbf{3.17} & \textbf{83.46} & \textbf{12.16} \\
\multicolumn{1}{c|}{w/o Interaction}        & 3.08          & 81.62          & 11.32            \\
\multicolumn{1}{c|}{w/o TF-CAF}              & 3.01          & 79.19          & 11.32          \\
\multicolumn{1}{c|}{w/o Interaction\&TF-CAF} & 2.91          & 78.13          & 10.46          \\ \bottomrule
\end{tabular}
\label{table5}
\end{center}
\end{table}

\setlength{\tabcolsep}{6pt}  
\begin{table*}[t]
\caption{Result comparisons between the proposed DBHN-Net and baselines in terms of the NB-PESQ, ESTOI and SI-SDR on the test set of the WSJ0-SI84+DNS-Challenge dataset. Factory1 and babble noises are adopted.}
\begin{center}
\begin{tabular}{cccccccccccccc}
\toprule 
\multicolumn{2}{c|}{\textbf{METRICS}}                                                                              & \multicolumn{4}{c|}{\textbf{PESQ}}                                                                                                                  & \multicolumn{4}{c|}{\textbf{ESTOI (in\%)}}                                                                                                                         & \multicolumn{4}{c}{\textbf{SI-SDR (in dB)}}                                                                                        \\ \midrule
\multicolumn{2}{c|}{\textbf{SNR (in dB)}}                                                                               & -5dB                          & 0dB                           & 5dB                           & \multicolumn{1}{c|}{AVG.}                        & -5dB                             & 0dB                              & 5dB                              & \multicolumn{1}{c|}{AVG.}                           & -5dB                          & 0dB                            & 5dB                            & AVG.                        \\ \midrule
\multicolumn{1}{c|}{}                            & \multicolumn{1}{c|}{Noisy}                              & 1.54                        & 1.86                        & 2.171                        & \multicolumn{1}{c|}{1.85}                        & 29.25                        & 43.11                        & 57.53                        & \multicolumn{1}{c|}{43.30}                        & -5.00                       & 0.00                         & 5.00                         & 0.00                           \\ \cmidrule{2-14} 
\multicolumn{1}{c|}{}                            & \multicolumn{13}{c}{ANN-based SE methods}                                                                               \\ \cmidrule{2-14} 
\multicolumn{1}{c|}{}                            & \multicolumn{1}{c|}{ConvTasNet$^{\cite{Conv-tasnet}}$}  & 2.11                        & 2.54                        & 2.88                        & \multicolumn{1}{c|}{2.52}                        & 60.06                       & 73.80                        & 82.90                         & \multicolumn{1}{c|}{72.25}                        & 6.56                        & 10.43                        & 13.63                        & 10.21                       \\
\multicolumn{1}{c|}{}                            & \multicolumn{1}{c|}{ DPRNN$^{\cite{DPRNN}}$}      & 2.17                        & 2.60                         & 2.96                        & \multicolumn{1}{c|}{2.57}                        & 61.74                        & 74.74                        & 83.53                        & \multicolumn{1}{c|}{73.34}                        & 6.88                       & 11.23                        & 13.82                        & 10.43                       \\
\multicolumn{1}{c|}{}                            & \multicolumn{1}{c|}{ LSTM$^{\cite{LSTM}}$}        & 1.97                        & 2.37                        & 2.67                        & \multicolumn{1}{c|}{2.34}                        & 49.33                        & 64.14                        & 74.98                       & \multicolumn{1}{c|}{62.82}                        & 2.49                        & 6.58                         & 9.54                         & 6.20                        \\
\multicolumn{1}{c|}{}                            & \multicolumn{1}{c|}{ CRN$^{\cite{CRN}}$}         & 1.97                        & 2.45                        & 2.79                        & \multicolumn{1}{c|}{2.41}                        & 50.52                        & 66.21                        & 77.24                        & \multicolumn{1}{c|}{64.66}                        & 2.66                        & 7.23                         & 10.79                        & 6.89                        \\
\multicolumn{1}{c|}{}                            & \multicolumn{1}{c|}{ GCRN$^{\cite{GCRN}}$}        & 2.02                        & 2.55                        & 2.92                        & \multicolumn{1}{c|}{2.50}                        & 56.44                        & 72.83                       & 82.08                        & \multicolumn{1}{c|}{70.45}                        & 5.36                        & 9.72                         & 12.67                        & 9.25                        \\
\multicolumn{1}{c|}{}                            & \multicolumn{1}{c|}{ DCCRN$^{\cite{dccrn}}$}       & 1.90                        & 2.46                        & 2.84                        & \multicolumn{1}{c|}{2.40}                        & 50.98                        & 68.06                        & 78.73                        & \multicolumn{1}{c|}{65.92}                        & 4.17                        & 8.61                         & 11.74                        & 8.17                        \\
\multicolumn{1}{c|}{}                            & \multicolumn{1}{c|}{FullSubNet$^{\cite{fullsubnet}}$}                         & 2.20                        & 2.64                        & 2.97                        & \multicolumn{1}{c|}{2.60}                        & 50.44                       & 67.34                        & 78.88                        & \multicolumn{1}{c|}{65.56}                        & 4.34                        & 9.01                         & 12.81                        & 8.72                        \\ 
\multicolumn{1}{c|}{}                            & \multicolumn{1}{c|}{CTS-Net$^{\cite{ctsnet}}$}                         & 2.32                        & 2.79                        & 3.14                        & \multicolumn{1}{c|}{2.75}                        & 62.92                       & 76.2                        & 84.35                        & \multicolumn{1}{c|}{74.49}                        & 6.75                        & 10.84                         & 13.96                        & 10.52                        \\
\multicolumn{1}{c|}{}                            & \multicolumn{1}{c|}{GaG-Net$^{\cite{gagnet}}$}                         & 2.36                        & 2.85                        & 3.22                        & \multicolumn{1}{c|}{2.81}                        & 65.84                       & 78.13                        & 85.79                        & \multicolumn{1}{c|}{76.59}                        & 7.36                        & 11.23                         & 14.31                        & 10.57                        \\
\multicolumn{1}{c|}{}                            & \multicolumn{1}{c|}{BSDB-Net$^{\cite{bsdb}}$}                         & 2.43                        & 2.92                        & 3.26                        & \multicolumn{1}{c|}{2.87}                        & 66.27                       & 78.19                        & 86.66                        & \multicolumn{1}{c|}{77.04}                        & 7.42                        & 11.34                         & 14.39                        & 11.05                        \\\cmidrule{2-14} 
\multicolumn{1}{c|}{}                            & \multicolumn{13}{c}{SNN-based SE methods}                                                                                                                    \\ \cmidrule{2-14} 
\multicolumn{1}{c|}{}                            & \multicolumn{1}{c|}{Spiking-UNet$^{\cite{u-netspiking}}$}                         & 1.96                        & 2.38                        & 2.98                        & \multicolumn{1}{c|}{2.44}                        & 47.13                        & 62.21                        & 76.11                        & \multicolumn{1}{c|}{61.82}                        & 2.78                        & 7.42                         & 9.87                         & 6.69                        \\ \cmidrule{2-14} 
\multicolumn{1}{c|}{}                            & \multicolumn{13}{c}{Proposed}                                                                                                                    \\ \cmidrule{2-14} 

\multicolumn{1}{c|}{\multirow{-16}{*}{ \rotatebox{90}{factory1 and babble} }} & \multicolumn{1}{c|}{\textbf{DBHN-Net}} & \textbf{2.72} & \textbf{3.17} &  \textbf{3.54} & \multicolumn{1}{c|}{ \textbf{3.14}} &  \textbf{71.21} &\textbf{83.46} & \textbf{89.29} & \multicolumn{1}{c|}{ \textbf{81.32}} & \textbf{8.47} &  \textbf{12.16} & \textbf{15.14} &  \textbf{11.93} \\ \bottomrule
\end{tabular}
\label{table6}
\end{center}
\vskip -0.1in
\end{table*}

\subsection{Loss Function and Training strategies}
\rev{Due to the non-differentiable nature of binary activation in SNN neurons, SNNs cannot directly perform backpropagation operation \cite{yudi}. Currently, the training of SNNs mainly involve two approaches. The first is to train the network using the ANN-SNN conversion strategy, and the second is to train the network by using gradient-proxy functions for backpropagation \cite{yudi}. The latter has shown promising results in recent years. In this paper, the Sigmoid function is leveraged as a gradient proxy function for SNN training, which can effectively handle the binary neuron outputs and enable gradient propagation during the backpropagation process. The gradient proxy function is defined as:}
\begin{equation}
\sigma(x)=\frac{1}{1+e^{-\alpha x}},
\end{equation}
\begin{equation}
{\sigma}' (x)=\alpha \cdot \sigma(x) \cdot (1-\sigma(x)).
\end{equation}

\rev{Among them, $\alpha$ is a hyperparameter used to adjust the gradient of the proxy function. The larger the value of $\alpha$ is, the greater the gradient of the function will be.}

The proposed dual-branch network is supervised by a combined ``RI+Mag'' loss to simultaneously optimize phase and magnitude estimations. In this formulation, RI refers to the complex spectrum, while Mag represents its magnitude.
\begin{equation}
\mathcal{L}_{RI}=\left \|\tilde{S}_{r}-S_r  \right \| ^2_F + \left \|\tilde{S}_{i}-S_i  \right \| ^2_F
\end{equation}
\begin{equation}
\mathcal{L}_{Mag}=\left \| \sqrt{|\tilde{S}_{r}|^2+|\tilde{S}_{i}|^2} + \sqrt{|{S}_{r}|^2+|{S}_{i}|^2}\right \|^2_{F} 
\end{equation}
\begin{equation}
\mathcal{L}=\beta \mathcal{L}_{RI}+(1-\beta)\mathcal{L}_{Mag}
\end{equation}
where $\left \| .\right \|_{F}$ represents Frobenius norm, and $\beta$ is empirically set to 0.5 \cite{bsdb}. Here, $\mathcal{L}{RI}$ and $\mathcal{L}{Mag}$ are the loss functions corresponding to the complex and magnitude spectra, respectively. The terms $S_r$ and $S_i$ refer to the real and imaginary components of the target complex spectrum. Their counterparts generated by the model are denoted as $\tilde{S}{r}$ and $\tilde{S}{i}$.

\subsection{Evaluation Metrics}
The enhancement performance was quantified using a comprehensive set of objective metrics. Speech quality was assessed via the Perceptual Evaluation of Speech Quality (PESQ) \cite{pesq} in both its narrow-band (NB-PESQ) and wide-band (WB-PESQ) versions. Intelligibility was measured using the Short-Time Objective Intelligibility (STOI) and its extended variant (ESTOI) \cite{estoi}. Furthermore, we reported the Scale-Invariant Signal Distortion Ratio (SISDR) to evaluate speech distortion \cite{mosSDR}, and a set of Mean Opinion Score (MOS) metrics—CSIG, CBAK, and COVL—as proxies for subjective perceptual quality \cite{mosSDR}.

\section{Results and Analysis}
\subsection{Ablation Study}
We conducted ablation studies on the WSJ0 dataset, examining four key aspects: (1) the effectiveness of the dual-path architecture, (2) the validity of using Mamba for sequence modeling in the ANN branch, (3) the efficacy of LIF-based residual structures in the SNN branch and the information refinement mechanism, and (4) the performance of the proposed Interaction module and TF-Cross Attention Fusion module for information fusion.
\subsubsection{Effect of Dual-Branch Model}
As shown in Table \ref{table2}, we systematically removed either the ANN or SNN branch to observe experimental effects. The results demonstrate that the dual-branch architecture significantly outperforms single-branch structures, with each branch playing crucial roles. The ANN branch contributes substantial performance improvements, while the SNN branch effectively reduces computational complexity and energy consumption. Notably, the Mamba-based sequence modeling in the ANN branch maintains performance while significantly reducing computational complexity.
\subsubsection{Effect of Mamba-Block}
As shown in Table \ref{table3}, in the proposed model, to ensure performance while compressing computational complexity as much as possible, we designed the Mamba Block as the sequence modeling module, which mainly includes T-Mamba and F-Mamba to model the data from different perspectives. Due to its unique linear complexity characteristics, Mamba has become a strong competitor to Transformer. Therefore, this ablation experiment aims to demonstrate the superiority of using Mamba for sequence modeling. According to the table, replacing Mamba with Transformer or LSTM significantly increases computational complexity while further reducing performance.
\subsubsection{Effect of SFEB and ITB}
In the SNN branch, to effectively mitigate the information loss caused by spike sequences (0/1), we designed the SFEB based on a residual structure and the ITB for further information refinement. As shown in the ablation experiments in Table \ref{table4}, we conducted ablation studies by replacing the SFEB with an LIF structure and directly removing the ITB structure. The experiments demonstrate that when the SFEB is replaced with LIF, performance decreases significantly, proving that our residual structure retains more critical information even when the input is binary (0/1). Meanwhile, removing the ITB also leads to a performance drop, indicating that the ITB module can further alleviate the inherent information loss in SNNs.

\setlength{\tabcolsep}{2.5pt}          
\begin{table}[]
\begin{center}
\caption{\rev{Result comparisons between the proposed DBHN-Net and baselines in terms of the WB-PESQ, STOI, CITG, CBAK, and COVL metrics on the VoiceBank+Demand dataset. “-” denotes the results are not reported in the original literature.}}
\begin{tabular}{ccccccc}
\toprule
\multicolumn{1}{c|}{\textbf{Model}}  & \multicolumn{1}{c|}{\textbf{Year}} & \textbf{WB-PESQ} & \textbf{STOI} & \textbf{CSIG} & \textbf{CBAK} & \textbf{COVL} \\ \midrule
\multicolumn{1}{c|}{noisy}  & \multicolumn{1}{c|}{—}             & 1.97             & 92.1          & 3.35          & 2.44          & 2.63          \\ \midrule
\multicolumn{7}{c}{\textbf{ANN-Speech Enhancement}}                                                                                                                  \\ \midrule
\multicolumn{1}{c|}{SEGA$^{\cite{segan}}$}           & \multicolumn{1}{c|}{2017}          & 2.16             & 92.5          & 3.48          & 29.4          & 2.8           \\
\multicolumn{1}{c|}{MMSEGAN$^{\cite{mmsegan}}$}         & \multicolumn{1}{c|}{2018}          & 2.53             & 93            & 3.8           & 3.12          & 3.14          \\
\multicolumn{1}{c|}{WaveNet$^{\cite{van2016wavenet}}$}         & \multicolumn{1}{c|}{2018}          & —               & —            & 3.62          & 3.32          & 2.98          \\
\multicolumn{1}{c|}{MetricGAN$^{\cite{metricgan}}$}       & \multicolumn{1}{c|}{2019}          & 2.86             &               & 3.99          & 3.18          & 3.42          \\
\multicolumn{1}{c|}{DCCRN$^{\cite{dccrn}}$}           & \multicolumn{1}{c|}{2020}          & 2.68             & 93.7          & 3.88          & 3.18          & 3.27          \\
\multicolumn{1}{c|}{PHASEN$^{\cite{Phasen}}$}          & \multicolumn{1}{c|}{2020}          & 2.99             & —            & 4.21          & 3.55          & 3.62          \\
\multicolumn{1}{c|}{MHSA-SPK$^{\cite{MHSA-SPK}}$}        & \multicolumn{1}{c|}{2020}          & 2.99             & —            & 4.15          & 3.42          & 3.53          \\
\multicolumn{1}{c|}{TSTNN$^{\cite{tstnn}}$}           & \multicolumn{1}{c|}{2021}          & 2.96             & 95            & 4.17          & 3.53          & 3.49          \\
\multicolumn{1}{c|}{CTS-Net$^{\cite{ctsnet}}$}         & \multicolumn{1}{c|}{2022}          & 2.92             & —            & 4.25          & 3.46          & 3.59          \\
\multicolumn{1}{c|}{GaG-Net$^{\cite{gagnet}}$}          & \multicolumn{1}{c|}{2022}          & 2.94             & 94.7          & 4.26          & 3.45          & 3.59          \\
\multicolumn{1}{c|}{SRTNET$^{\cite{srtnet}}$}          & \multicolumn{1}{c|}{2023}          & 2.69             & —            & 4.12          & 3.19          & 3.39          \\
\multicolumn{1}{c|}{CompNet$^{\cite{compnet}}$}         & \multicolumn{1}{c|}{2023}          & 2.90             & —            & 4.16          & 3.37          & 3.53          \\
\multicolumn{1}{c|}{FDFNet$^{\cite{FDFnet}}$}          & \multicolumn{1}{c|}{2024}          & 3.05             & —            & 4.23          & 3.55          & 3.65          \\
\multicolumn{1}{c|}{S4DSE$^{\cite{s4sd}}$}           & \multicolumn{1}{c|}{2024}          & 2.55             & —            & 3.94          & 3.00          & 3.32          \\
\multicolumn{1}{c|}{BSDBNet$^{\cite{bsdb}}$}         & \multicolumn{1}{c|}{2025}          & 3.17             & 94.8          & \textbf{4.32}          & 3.58          & 3.71          \\ \midrule
\multicolumn{7}{c}{\textbf{SNN-Speech Enhancement}}                                                                                                                                                                                                                                                    \\ \midrule
\multicolumn{1}{c|}{Spiking-UNet$^{\cite{u-netspiking}}$}   & \multicolumn{1}{c|}{2023}          & 2.66             & 92            & —            & —            & —            \\
\multicolumn{1}{c|}{Fullsub-Spiking$^{\cite{Spiking-FullSubNet}}$} & \multicolumn{1}{c|}{2024}          & 2.79             & 93.7          & 3.96          & 3.26          & 3.29          \\ \midrule
\multicolumn{7}{c}{\textbf{Proposed}}                                                                                                                                                                                                    \\ \midrule
\multicolumn{1}{c|}{\textbf{DBHN-Net}} & \multicolumn{1}{c|}{2025}                 & \textbf{3.08}    & \textbf{95}   & \textbf{4.32} & \textbf{3.59} & \textbf{3.74} \\ \bottomrule
\end{tabular}
\label{table7}
\end{center}
\end{table}

\subsubsection{Effect of TF-CAF and Interaction module}
In the proposed model, the SNN branch is employed to reduce power consumption due to its characteristic of propagating binary information, while the ANN branch better extracts data information to mitigate the impact of information loss. Our model constructs an Interaction module and a TF-CAF module to achieve comprehensive information fusion across all stages. As shown in the ablation study in Table \ref{table5}, when both modules are removed simultaneously, the model performance drops significantly, demonstrating that sufficient information fusion is crucial and that our modules perform effectively. Furthermore, we observe that the TF-CAF module contributes more substantially, indicating that the cross-attention fusion mechanism divided into time and time-frequency domains aligns well with the characteristics of speech information, yielding highly favorable results.
\subsubsection{Visualization experiment on ablation study}
\begin{figure*}[!t]
\begin{center}
\centerline{\includegraphics[width=0.8\textwidth]{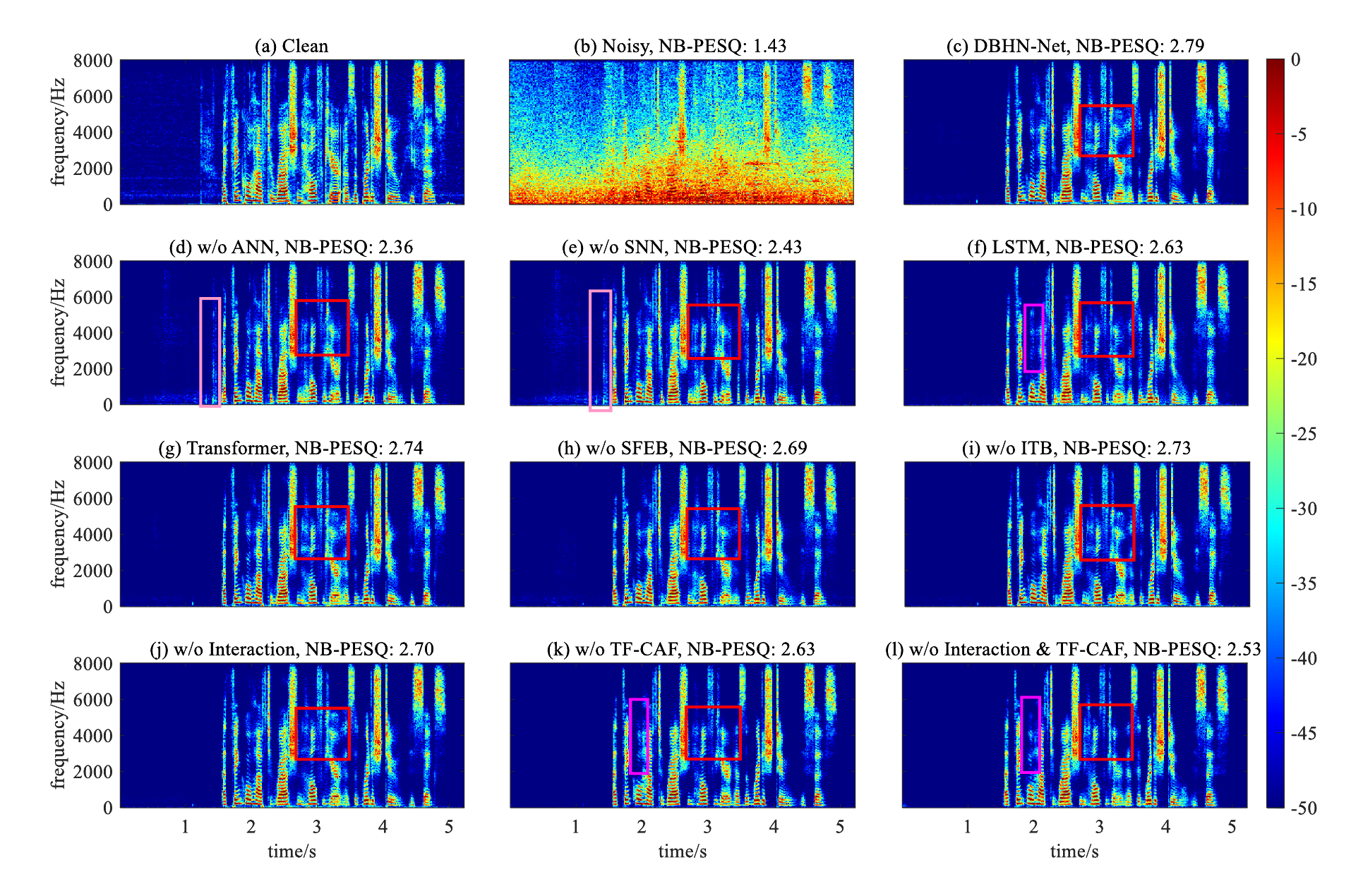}}
\caption{The figure shows the visualization results of ablation experiments. The models were trained on the WSJ0-SI84 and DNS-Challenge datasets, and tested under the condition of -5 dB signal-to-noise ratio (SNR) using highly non-stationary noise from NOISEX92.}
\label{fig5}
\end{center}
\end{figure*}

As shown in Figure \ref{fig5}, we conducted visualization experiments on relevant ablation studies under extremely low signal-to-noise ratios (e.g. -5dB). First, it can be clearly observed that using only a single branch (as shown in Figures \ref{fig5}.d and \ref{fig5}.e) exhibits spectrogram distortion (red boxes) and residual noise (pink boxes) on the spectrogram. This indicates that the single-branch architecture, particularly the pure SNN branch, leads to a significant performance degradation. Second, replacing the Mamba\_block modules in our network with corresponding Transformer and LSTM modules(as shown in Figures \ref{fig5}.f and \ref{fig5}.g) for sequence modeling. Compared to our DBHN-Net, these replacements cause noticeable spectrogram distortion (The red box indicates spectrogram distortion) . This demonstrates that our designed Mamba\_block effectively maintains performance while reducing computational requirements. Furthermore, ablation experiments on the SNN branch reveal that replacing SEFB with LIF or removing ITB (as shown in Figures \ref{fig5}.h and \ref{fig5}.i) both lead to varying degrees of spectral detail loss (as highlighted in the pink and red boxes) . This confirms that the residual structures specifically designed for spiking information effectively preserve critical features. Finally, for the information interaction modules, removing either the Interaction block or the final TF-CAF module (as shown in Figures \ref{fig5}.j, \ref{fig5}.k and \ref{fig5}.l) causes significant performance drops. Particularly, the TF-CAF module plays a significant role in the restoration of spectrogram details, indicating that effective and thorough interaction between the two branches is essential. And comprehensive information fusion leads to superior performance.

\subsection{Comparisons with Baselines on WSJ0+DNS-Challenge Datasets}
On the WSJ0-84+DNS-Challenge dataset, objective evaluations using PESQ, ESTOI, and SI-SDR demonstrate that our proposed model outperforms all baseline methods. We selected highly non-stationary noises, specifically Factory1 and Babble, as the test set noises. As shown in Table \ref{table6}, we compared against both ANN-based and SNN-based baseline models. In particular, Spiking-UNet is the only SNN-based model that has been experimentally validated on this dataset. For fair comparison, all reported results are directly taken from the original publications. First, the proposed model achieves SOTA performance among both ANN-based and SNN-based speech enhancement models, demonstrating that our fusion strategy effectively mitigates information loss caused by spiking signals while maintaining low computational complexity. Second, compared to parallel dual-branch models such as CTS-Net and GaG-Net, our approach places greater emphasis on fusion strategies by implementing information interaction across all stages of the model and incorporating cross-attention fusion at the final stage. This indicates that our model can capture critical information more effectively from each branch. Finally, while baseline models often exhibit limited enhancement capability under low SNR conditions, our model maintains robust performance even in extremely low SNR scenarios. These results suggest that appropriately discarding redundant information under severe noise contamination can lead to better enhancement outcomes. Collectively, this study demonstrates the advantage of fusion networks in challenging environments (low SNR, highly non-stationary noise), achieving an optimal balance between computational efficiency and performance while effectively extracting and modeling semantically important information .

\setlength{\tabcolsep}{2.5pt}     
\begin{table}[]
\begin{center}
\caption{\rev{Result comparisons between the proposed DBHN-Net and baselines in terms of the WB-PESQ, NB-PESQ, STOI, SI-SDR metrics on the DNS-Challenge dataset. “-” denotes the results are not reported in the original literature.}}
\begin{tabular}{c|ccccc}
\toprule
\textbf{Model}          & \textbf{Years} & \textbf{WB-PESQ} & \textbf{PESQ} & \textbf{STOI} & \textbf{SI-SDR} \\ \midrule
Noisy                   & —             & 1.58             & 2.45             & 91.52             & 9.07            \\
NSNet$^{\cite{inteldnschallenge}}$                   & 2020           & 2.15             & 2.87             & 94.47             & 15.61           \\
DTLN$^{\cite{dtln}}$                   & 2020           & 2.34             & 3.04             & 95                & 16.34           \\
DCCRN$^{\cite{dccrn}}$                   & 2020           & —               & 3.27             & -                & —              \\
FullSubNet$^{\cite{fullsubnet}}$              & 2021           & 2.78             & 3.31             & 96.11             & 17.29           \\
DCCRN+$^{\cite{dccrn+}}$                  & 2021           & —               & 3.33             & —                & —              \\
CTS-Net$^{\cite{ctsnet}}$                 & 2022           & 2.94             & 3.42             & 96.66             & 17.99           \\
GaG-Net$^{\cite{gagnet}}$                  & 2022           & 3.18             & 3.57             & 93.22             & 16.57           \\
TalorSENet$^{\cite{li2022taylor}}$              & 2022           & 3.22             & 3.59             & 98                & 19.15           \\
IIFC-Net$^{\cite{IIFC-Net}}$                & 2023           & 3.37             & —               & 98                & 19.94           \\
MFNet$^{\cite{MFNet}}$                   & 2023           & 3.43             & \textbf{3.74}             & 98                & 20.31           \\
SICRN$^{\cite{SICRN}}$                   & 2024           & 2.62             & 3.23             & 96                & 15.6            \\
CMGAN-Bispectra$^{\cite{CMGANB}}$         & 2025           & 3.35             & —               & 98                & 14.72           \\
TSDT-Net$^{\cite{TSDT-Net}}$                & 2025           & 2.68             & —               & —                & 19.55           \\ \midrule
\textbf{DBHN-Net} & \textbf{2025}  & \textbf{3.45}    & \textbf{3.74}    & \textbf{98}       & \textbf{20.63}  \\ \bottomrule
\end{tabular}
\label{table8}
\end{center}
\end{table}
\subsection{Comparisons with Baselines on VoiceBank+Demand Datasets}
We evaluated the proposed model on another public dataset, VoiceBank + Demand, and compared it with various baseline models. We additionally included an SNN-based model, Fullsub-Spiking, which won the Intel Neuromorphic DNS Challenge. As shown in Table \ref{table7} , several key observations can be made: First, SNN-based speech enhancement models generally demonstrate inferior performance compared to ANN-based models, with a significant performance gap between them. This indicates that converting complex speech signals into spike sequences causes substantial information loss. However, our fusion strategy effectively mitigates this information loss and achieves remarkably superior results. Second, while significantly reducing computational complexity compared to ANN-based models, our approach maintains competitive performance. This suggests that certain architectural components in ANN-based models contribute to increased computational overhead without providing substantial practical benefits. The experimental results on this public dataset demonstrate that our proposed model successfully addresses two critical challenges: it resolves the high energy consumption problem inherent in ANN-based models while simultaneously overcoming the information loss issue associated with SNN-based approaches. Ultimately, by leveraging the respective advantages of both architectures, our model achieves comprehensive improvements across all performance dimensions.

\begin{table}[]
\setlength{\tabcolsep}{18pt}
\begin{center}
\caption{An Ablation Study Focusing on the Computational Complexity of Different Model Variants.}
\begin{tabular}{ccccc}
\toprule
\multicolumn{1}{c|}{\textbf{Model}}                      & \textbf{MACs}    \\ \midrule
\multicolumn{1}{c|}{LSTM$^{\cite{LSTM}}$}                          & 3.69 G/s         \\
\multicolumn{1}{c|}{CRN$^{\cite{CRN}}$}                           & 2.54 G/s         \\
\multicolumn{1}{c|}{GCRN$^{\cite{GCRN}}$}                          & 2.40 G/s         \\
\multicolumn{1}{c|}{FullSubNet$^{\cite{fullsubnet}}$}                   & 29.83 G/s        \\
\multicolumn{1}{c|}{CTSNet$^{\cite{ctsnet}}$}                      & 5.48 G/s         \\
\multicolumn{1}{c|}{ConvTasNet$^{\cite{Conv-tasnet}}$}                   & 5.22 G/s         \\
\multicolumn{1}{c|}{DPRNN$^{\cite{DPRNN}}$}                        & 8.47 G/s         \\
\multicolumn{1}{c|}{DDAEC$^{\cite{ddeac}}$}                        & 36.85 G/s        \\
\multicolumn{1}{c|}{GaG-Net$^{\cite{gagnet}}$}                     & 2.81 G/s         \\
\multicolumn{1}{c|}{BSDBNet$^{\cite{bsdb}}$}                      & 1.68 G/s         \\ \midrule
\multicolumn{1}{c|}{\textbf{DBHN-Net}}   & \textbf{1.32G/s} \\ \bottomrule
\end{tabular}
\label{table9}
\end{center}
\end{table}

\subsection{Comparisons with Baselines on DNS-Challenge2020 Dataset}
The objective results on the DNS-Challenge dataset, summarized in Table \ref{table8}, demonstrate that our proposed model achieves superior performance across key metrics—WB-PESQ, NB-PESQ, STOI, and SI-SDR—when benchmarked against existing methods. All baseline results are directly sourced from their original publications. Taking TSDT-Net as a representative example, our proposed model demonstrates significant improvements of approximately 0.76 and 0.96 in WB-PESQ and SI-SDR metrics, respectively. This substantial enhancement clearly indicates the effectiveness of our algorithm in noise suppression tasks. Furthermore, while current SNN-based speech enhancement models predominantly employ multi-stage architectures (e.g., CTS-Net, GaG-Net), which utilizing Transformer or LSTM modules for intermediate sequence modeling and resulting in considerable structural complexity our approach presents a streamlined alternative. By integrating a Mamba-based ANN branch for sequence modeling with an SNN branch for efficient feature extraction, our model not only reduces architectural complexity but also achieves superior performance. Notably, the DNS-Challenge dataset contains multilingual audio content including French, English, and Indian languages, presenting a more rigorous test of cross-lingual robustness compared to other datasets. The consistent performance gains achieved by our proposed model across all evaluation metrics demonstrate its exceptional generalization capability in diverse linguistic environments.

\subsection{Model Complexity Comparison}
Our model delivers a substantial efficiency gain, as evidenced by the computational complexity analysis in Table \ref{table9}. When evaluated on standardized one-second audio clips from the WSJ0-SI84 + DNS2020 dataset, it reduces operations by an average of 7.5× relative to the baselines, all while delivering comparable enhancement performance. This significant complexity reduction clearly indicates that introducing an SNN branch into ANN-based dual-branch architectures can substantially decrease computational requirements. Specifically, the ANN branch of our model employs Mamba-based sequence modeling, which demonstrates excellent effectiveness with linear complexity. Simultaneously, the residual structure in the SNN branch effectively mitigates the impact of information loss during spike encoding. Furthermore, the Interaction module enables efficient information exchange across all stages of the model, while the cross-attention fusion module further reduces information loss during spike conversion in a data-driven manner. Collectively, these innovations allow the proposed model to achieve an optimal balance between computational efficiency and enhancement performance.

\section{Conclusion}
In this work, we propose a Dual-Branch Hybrid Neural Network for single-channel speech enhancement, which integrates both ANN and SNN architectures. Specifically, we design a dual-branch parallel network consisting of an ANN branch and an SNN branch. In the ANN branch, we incorporate a Band-Split module and a Mamba module to reduce computational complexity while maintaining performance. In the SNN branch, we implement residual-structured LIF units and an Information Transformation Block to mitigate information loss caused by the conversion of floating-point data into spike sequences. Additionally, we design an Interaction module and a TF-Cross Attention Fusion module to facilitate inter-branch information exchange and final feature fusion. Extensive ablation studies on the network architecture and comprehensive comparisons with baseline models on three public datasets demonstrate the superior performance of our proposed approach across all evaluation metrics. For future work, we will further explore the role of SNNs in reducing computational energy consumption and investigate more effective solutions to address information loss. We also plan to extend this framework to related tasks such as multi-channel speech enhancement, dereverberation, and target speaker extraction. Furthermore, as the utilization of SNNs remains relatively underdeveloped, exploring optimal strategies for integrating SNNs with ANNs presents a challenging yet promising research direction.
\bibliographystyle{IEEEtran}
\normalem
\small
\bibliography{IEEEabrv, bibfile.bib}

\vspace{-10mm}
\begin{IEEEbiography}
[{\includegraphics[width=1in,height=1.15in,clip,keepaspectratio]{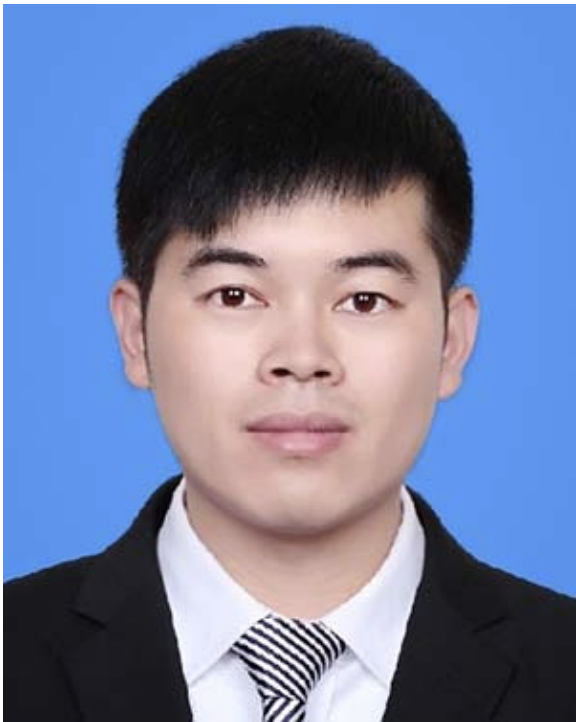}}]
	{Cunhang Fan} 
    (Member, IEEE) received the BS degree from the Beijing University of Chemical Technology (BUCT), Beijing, China, in 2016 and the PhD degree with the National Laboratory of Pattern Recognition (NLPR), Institute of Automation, Chinese Academy of Sciences (CASIA), Beijing, China, in 2021. He is currently an associate professor with the School of Computer Science and Technology, Anhui University, Heifei, China. His current research interests include speech enhancement, fake speech detection, speech recognition and speech processing.
\end{IEEEbiography}

\vspace{-10mm}
\begin{IEEEbiography}
[{\includegraphics[width=1in,height=1.15in,clip,keepaspectratio]{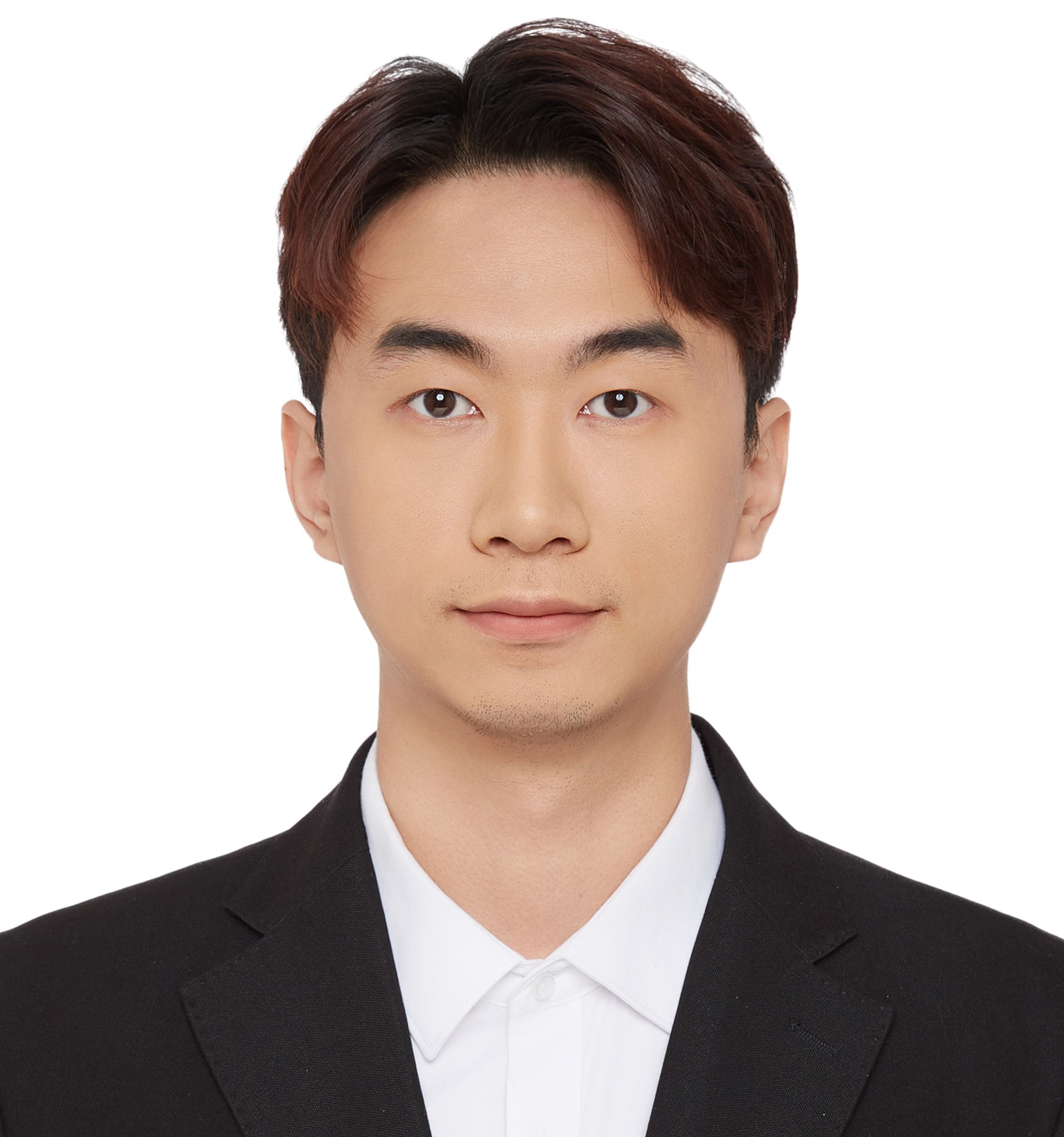}}]
	{Enrui Liu} 
	received the B.Sc. degree from Shandong University of Finance and Economics, Shandong, China, in 2019. He is currently working toward the M.Sc. degree with Anhui University, Anhui, China. His research interests include speech enhancement and spiking neural networks, with particular focus on reducing complexity and power consumption while maintaining performance in single-channel speech enhancement algorithms.
\end{IEEEbiography}

\vspace{-10mm}
\begin{IEEEbiography}
[{\includegraphics[width=1in,height=1.15in,clip,keepaspectratio]{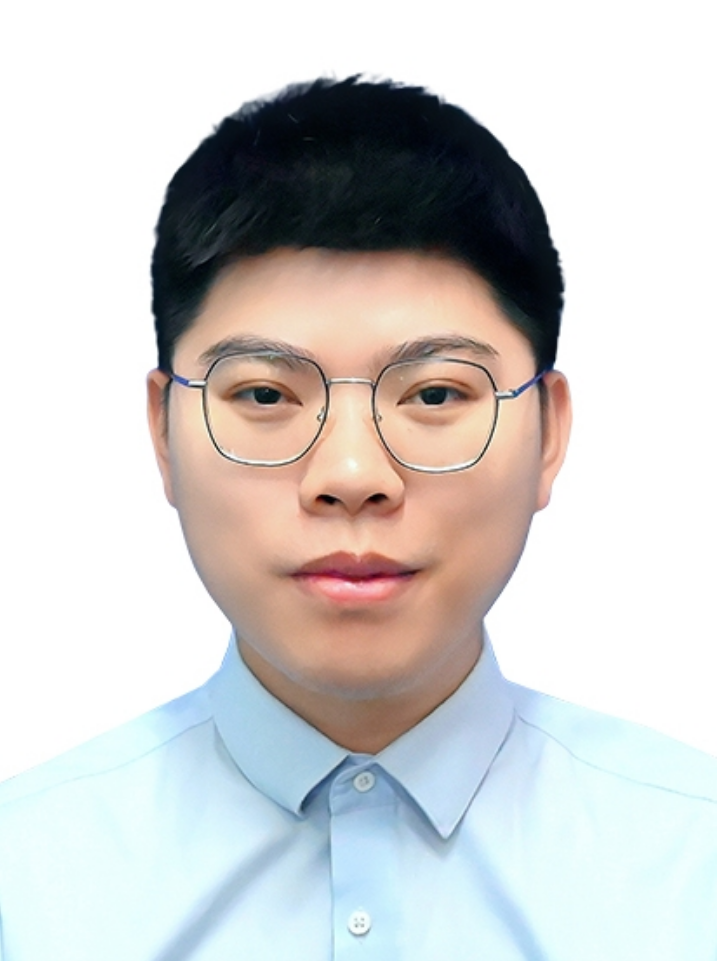}}]
	{Jing Zhou} 
	received the B.S. degree in Electronic Information Engineering and the M.S. degree in Electronics and Communication Engineering from Jiangxi University of Science and Technology, China, in 2017 and 2020, respectively. In 2024, he received the Ph.D. degree in Electronic Science and Technology from Beijing University of Technology, China. Since 2024, he has been working as a Speech Algorithm Engineer at China Telecom Artificial Intelligence Technology (Beijing) Co., Ltd. His research interests include speech enhancement, speech separation, microphone-array signal processing, and machine learning.
\end{IEEEbiography}
\vspace{-10mm}
\begin{IEEEbiography}
[{\includegraphics[width=1in,height=1.15in,clip,keepaspectratio]{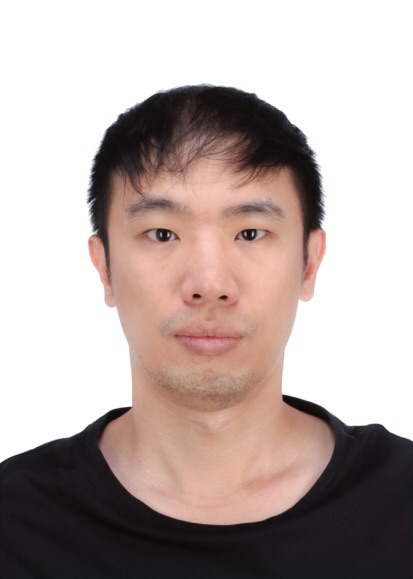}}]
	{Jian Kang} 
	received his B.S. degree and Ph.D. degree from the Department of Electronic Engineering at Tsinghua University in 2013 and 2018, respectively. His main research areas include large-scale speech models, speech recognition, speech translation and speech language model.
\end{IEEEbiography}
\vspace{-10mm}
\begin{IEEEbiography}
[{\includegraphics[width=1in,height=1.15in,clip,keepaspectratio]{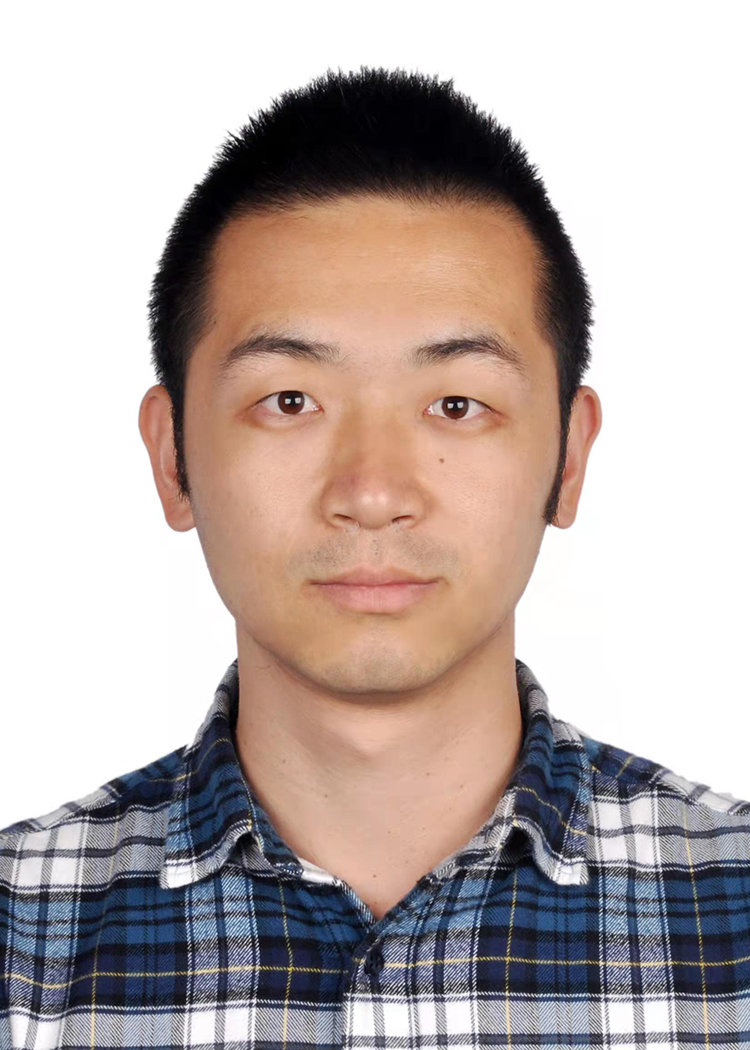}}]
	{Jie Li} 
	received his B.E. degree in Automation from Tianjin University, China, in 2011, and his Ph.D. degree in Pattern Recognition and Intelligent Systems from the Institute of Automation, Chinese Academy of Sciences (CASIA), China, in 2016. His doctoral research focused on intelligent speech information processing. In 2022, he joined China Telecom Artificial Intelligence Technology (Beijing) Co., Ltd. as a Speech Algorithm Engineer. His current research interests include speech signal processing, speech recognition, speech synthesis, and speech foundation models.
\end{IEEEbiography}
\vspace{-10mm}
\begin{IEEEbiography}
[{\includegraphics[width=1in,height=1.15in,clip,keepaspectratio]{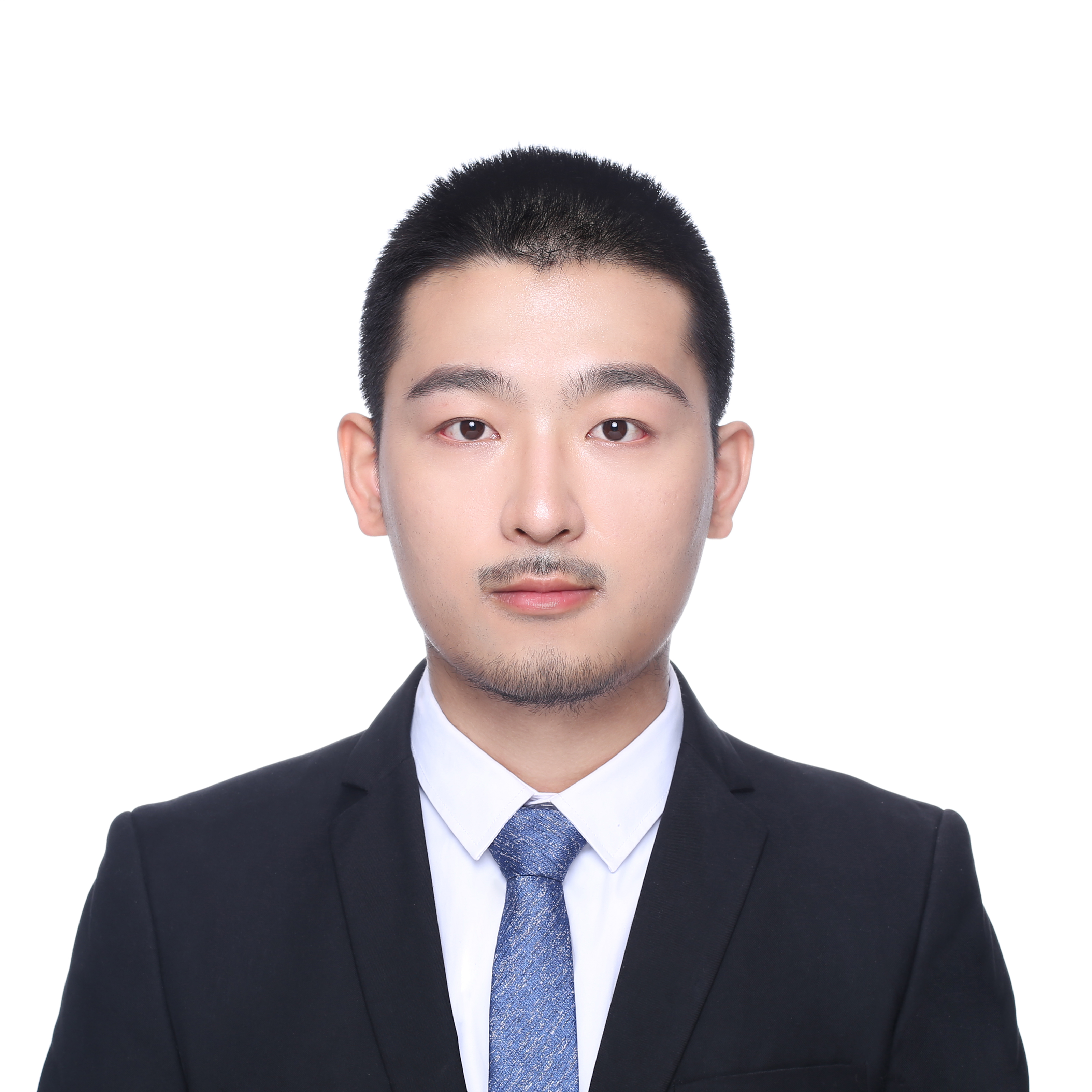}}]
	{Andong Li} 
	(Member, IEEE) received the B.S. degree in information engineering from Southeast University, Nanjing, China, in 2018, and the Ph.D. degree in signal and information processing from the Institute of Acoustics, Chinese Academy of Sciences, Beijing, China, in 2023. From 2023 to 2024, he was a Senior Researcher with Tencent AI Lab. He is currently an Associate Researcher with the Institute of Acoustics, Chinese Academy of Sciences. His research interests include speech enhancement, audio coding, array signal processing, and speech editing. He is also an active reviewer for multiple leading conferences and journals, such as INTERSPEECH, ICASSP, AAAI, ICLR, IJCAI, ACM MM, IEEE Signal Processing Letters, Speech Communication, Neural Networks, Pattern Recognition, and IEEE/ACM Transactions on Audio, Speech, and Language Processing.
\end{IEEEbiography}
\vspace{-10mm}
\begin{IEEEbiography}
[{\includegraphics[width=1in,height=1.15in,clip,keepaspectratio]{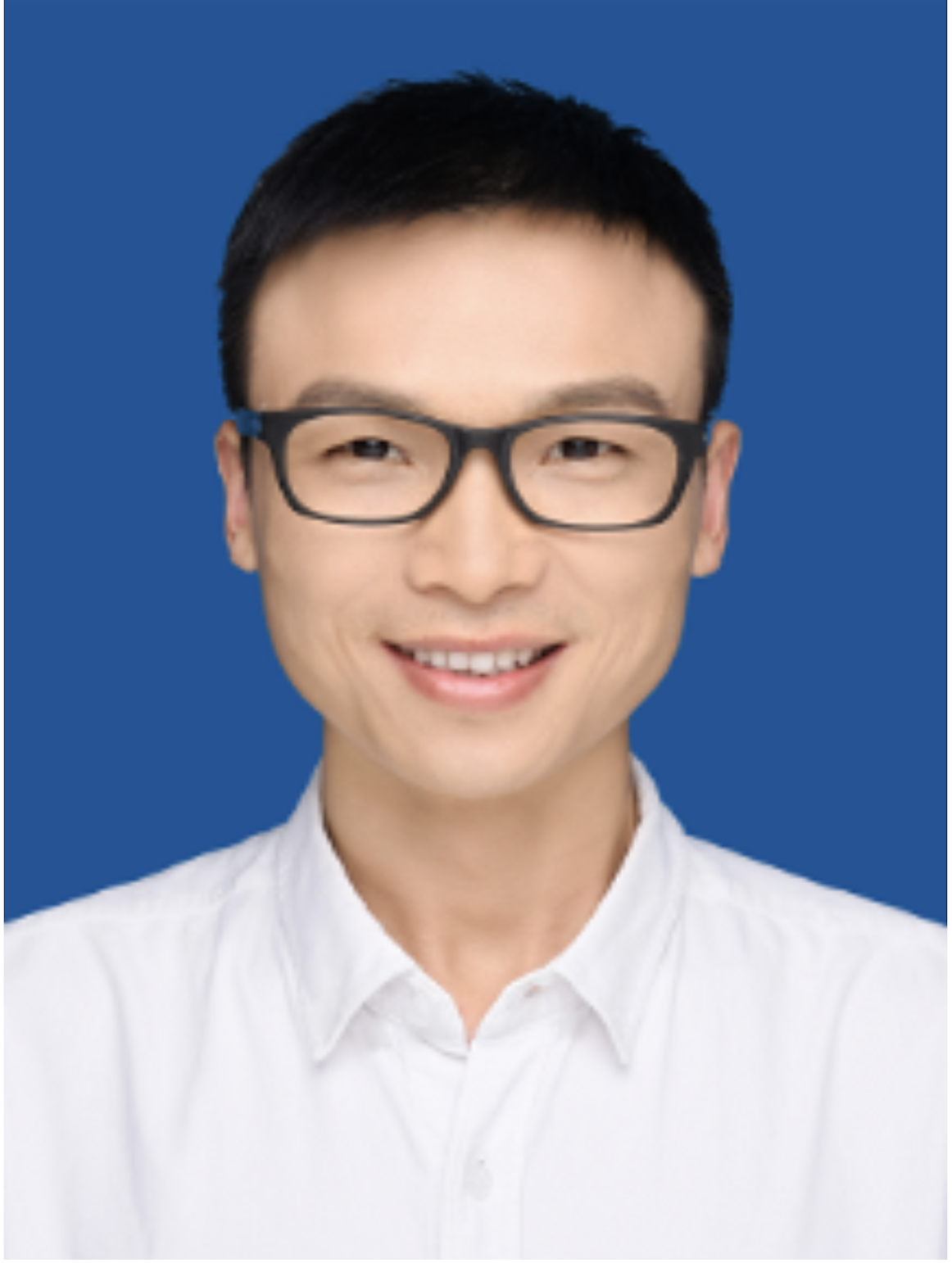}}]
	{Jian Zhou} 
	received the B.S. and M.S. degrees from Southwest Jiaotong University, Chengdu, China, in 2004 and 2006, respectively, and the Ph.D. degree from Southeast University, Nanjing, China, in 2013. He is currently an Associate Professor with Anhui University, Anhui, China. His research interests include artificial intelligence, pattern recognition, multimedia information processing, speech recognition, and voice conversion.
\end{IEEEbiography}
\vspace{-10mm}
\begin{IEEEbiography}
[{\includegraphics[width=1in,height=1.15in,clip,keepaspectratio]{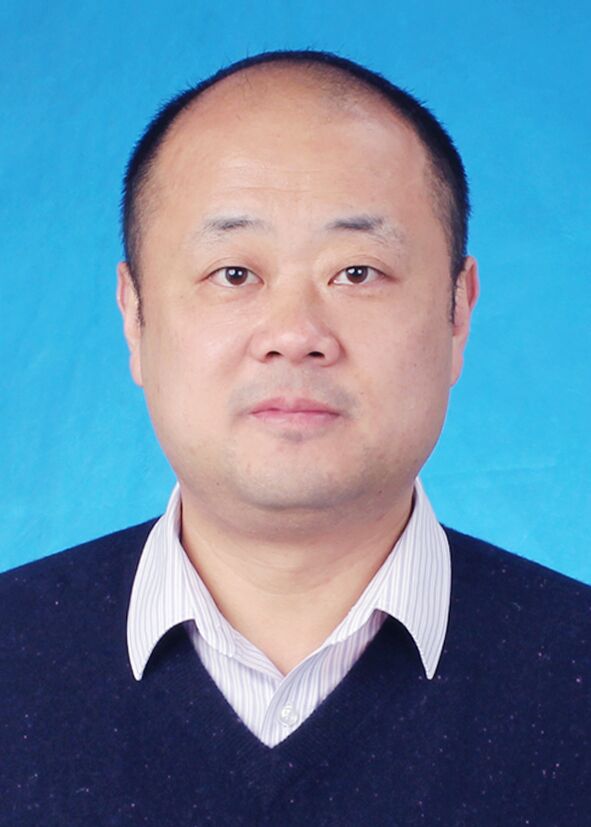}}]
	{Zhao Lv} 
	(Member, IEEE) received the PhD degree in computer application technology from Anhui University, Hefei, China, in 2011. He was a visiting scholar with the University of Utah, Salt Lake City, USA, from 2017 to 2018. He is currently a professor with the School of Computer Science and Technology, Anhui University, Hefei, China. His research interests include intelligent information processing and pattern recognition regarding biomedical signals (EEG, EOG, etc.) as well as speech signal processing.
\end{IEEEbiography}
\vspace{-10mm}
\begin{IEEEbiography}
[{\includegraphics[width=1in,height=1.15in,clip,keepaspectratio]{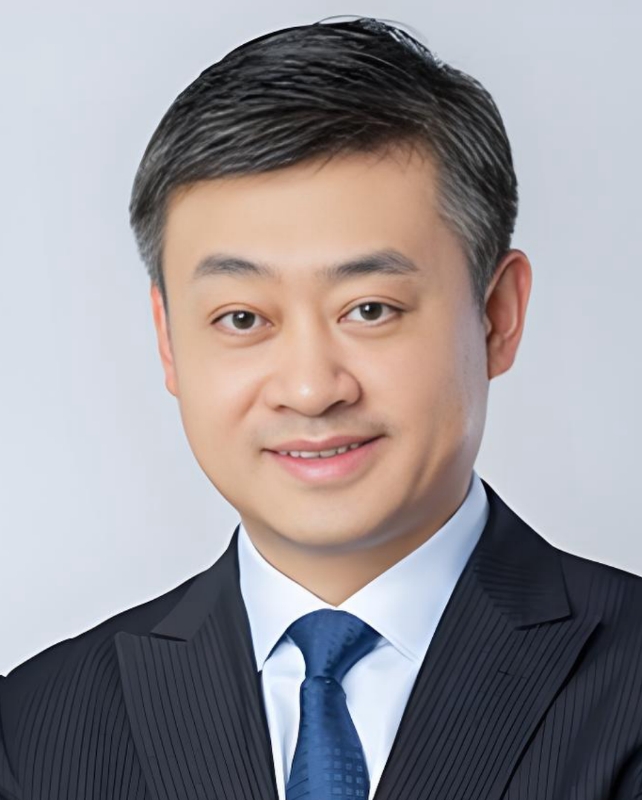}}]
	{Xuelong Li} 
	(Fellow, IEEE) is the CTO and chief scientist with China Telecom, where he founded the Institute of Artificial Intelligence (TeleAl) of China Telecom.
\end{IEEEbiography}
\end{document}